\documentclass[11pt, aps, notitlepage,superscriptaddress,nofootinbib]{revtex4-1}
\usepackage{comment}
\usepackage{amsmath}
\usepackage{amssymb}
\usepackage{amsthm}
\usepackage{graphicx}
\usepackage{subcaption}
\usepackage{float}
\usepackage[justification=raggedright, singlelinecheck=off]{caption}
\usepackage{tikz}
\usetikzlibrary{shapes,arrows,decorations.markings}

\newcommand{\ket}[1]{\left|#1\right\rangle}
\newcommand{\bra}[1]{\left\langle#1\right|}

\begin{document}
\title{Adiabatic Quantum Simulation of Quantum Chemistry}

\author{Ryan Babbush}
\affiliation{Department of Chemistry and Chemical Biology, Harvard University, Cambridge, MA 02138 USA}
\author{Peter J. Love}
\affiliation{Department of Physics, Haverford College, Haverford, PA 19041, USA}
\author{Al\'{a}n Aspuru-Guzik}
\email{Corresponding author Email: aspuru@chemistry.harvard.edu}
\affiliation{Department of Chemistry and Chemical Biology, Harvard University, Cambridge, MA 02138 USA}
\date{\today}

\begin{abstract}
We show how to apply the quantum adiabatic algorithm directly to the quantum computation of molecular properties. We describe a procedure to map electronic structure Hamiltonians to $2$-local qubit Hamiltonians with a small set of physically realizable couplings. By combining the Bravyi-Kitaev construction to map fermions to qubits with perturbative gadgets to reduce the Hamiltonian to $2$-local, we obtain precision requirements on the coupling strengths and a number of ancilla qubits that scale polynomially in the problem size. Hence our mapping is efficient. The required set of controllable interactions includes only two types of interaction beyond the Ising interactions required to apply the quantum adiabatic algorithm to combinatorial optimization problems. Our mapping may also be of interest to chemists directly as it defines a dictionary from electronic structure to spin Hamiltonians with physical interactions.
\end{abstract}
\maketitle

\section{Introduction}

Adiabatic quantum computing (AQC) works by changing the Hamiltonian of a controllable quantum system from an initial Hamiltonian whose ground state is easy to prepare into a Hamiltonian whose ground state encodes the solution of a computationally interesting problem~\cite{Farhi2000,Farhi2001}. The speed of this algorithm is determined by the adiabatic theorem of quantum mechanics which states that an eigenstate remains at the same position in the eigenspectrum if a perturbation acts on the system sufficiently slowly \cite{Born,Farhi2000,Boixo2009}. Simply embedding a computational problem in a Hamiltonian suitable for AQC does not ensure an efficient solution. The required runtime for the adiabatic evolution depends on the energy gap between the ground state and first excited state at the smallest avoided crossing~\cite{Farhi2000}.

AQC has been applied to classical optimization problems that lie in the complexity class NP~\cite{GarveyJohnson}.  For example, studies have been performed on satisfiability~\cite{Hogg2003,Choi2010,Nehaus2011}, Exact Cover~\cite{Farhi2000,Farhi2001}, 3-regular 3-\textsc{XORSAT} and 3-regular Max-Cut~\cite{Farhi2012}, random instances of classical Ising spin glasses~\cite{Boixo2013}, protein folding~\cite{Perdomo-Ortiz2012,Babbush2012} and machine learning~\cite{Denchev2012,Neven2009}. AQC has also been applied to structured and unstructured search~\cite{Roland:2002,Roland:2003}, and even search engine ranking~\cite{Garnerone2012} and artificial intelligence problems arising in space exploration~\cite{Smelyanskiy2012}. Many of these applications follow naturally from the NP-Completeness of determining the ground state energy of classical Ising spin glasses~\cite{Barahona1982}. This creates an equivalence between a very large set of computational problems (the class NP) and a set of models in classical physics (classical Ising models with random coupling strengths). The advent of AQC provides a powerful motivation to study the detailed implications of this mapping. In general, we do not expect that quantum computing, including AQC, can provide efficient solutions to NP-Complete problems {\em in the worst case}~\cite{Bernstein:1997}. However, there may exist sets of instances\footnote{The word problem is used in computer science to mean a type of problem. For example, the problem \textsc{CROSSWORD} can be defined as finding the solution to any imaginable crossword puzzle. An instance of \textsc{CROSSWORD} would be one particular crossword puzzle. Different sets of instances can have different difficulties of solution, just as the difficulty of the New York Times crossword puzzle varies with the day of the week.} of some NP-Complete problems for which AQC can find the ground state efficiently, but which defy efficient classical solution by any means. If this is the case then AQC is certainly of considerable scientific interest, and likely of great industrial importance. 

The potential value of a positive answer to this conjecture has motivated a commercial effort to construct an adiabatic quantum computer~\cite{Harris2007,Harris2008,Harris2009b,Harris2009c,Lanting2009,Johansson2009,Berkley2010}. Currently, these experimental implementations of AQC are not strictly confined to the ground state at zero temperature but have considerable thermal mixing of higher lying states. Such intermediate implementations are referred to as quantum annealing devices. Quantum annealing machines with up to $509$ qubits have been commercially manufactured by \emph{D-Wave Systems} \cite{Berkley2013,Johnson2011,Dickson2013}. They are currently the subject of serious scientific investigation to determine whether their operation depends significantly on their quantum properties, and if so, whether it provides a speedup for any class of instances~\cite{Johnson2011,Pudenz2013,Bian2013,Boixo2013,Wang2013,Smolin2013,Boixo2013}. 

Quantum computers have been rigorously proved to provide an algorithmic advantage over the best known classical approaches\footnote{In many cases, including factoring, the minimal classical cost is not known.} for a small set of problems~\cite{Shor:1997,Childs:2003,Grover:1996}. Adiabatic quantum computation applied to classical Ising Hamiltonians (equivalently, all problems in NP) gives an approach to a very large class of problems where the advantage (if any) is currently unknown. The construction of medium scale ($500$ qubit) quantum annealing machines provides a hardware platform where the properties of AQC can be investigated experimentally. Such investigations have already been performed for many problems. At present, optimized codes on classical hardware can find the ground state of many instances in comparable time to the D-Wave device \cite{Boixo2013}.  However, even if no interesting set of instances is found on which quantum annealing on the classical Ising model outperforms classical approaches, the hardware constructed to date represents an important step towards the construction of large scale quantum information technology. If we regard quantum annealing of the classical Ising model as the first step - what is the natural next step to take? 

Quantum simulation has provided a rich set of questions and methods in quantum computation since Feynman's suggestion that quantum devices would be best suited to computation of quantum properties~\cite{Feynman1982}. This observation has been fleshed out through early work on specific systems~\cite{Meyer1996,Wiesner1996,Lloyd1996,Lidar1997,Boghosian1998,Zalka1998} and through quantum algorithms for computation of eigenvalues, dynamics and other properties~\cite{Abrams1999,Berry2007,Kassal2008,Wiebe2008,Ward2009,Sanders2012,Sanders2013}. Recently, there have been many proposals for the simulation of quantum lattice models using trapped ions, trapped atoms and photonic systems~\cite{Weimer2010,Ma2011,Hague2013,Cohen2013,Hauke2013}. There has been rapid experimental progress in the quantum simulation of a number of systems~\cite{Simon2011,Greiner2009,wineland2002,Schaetz2008,Johanning2009,Ma2012b,Monroe2013}. A natural target for these simulations is the phase diagram of the Fermi-Hubbard model - believed to inform our understanding of high-$T_c$ superconductivity. For this reason many of these approaches are aimed at simulating systems of interacting fermions. 

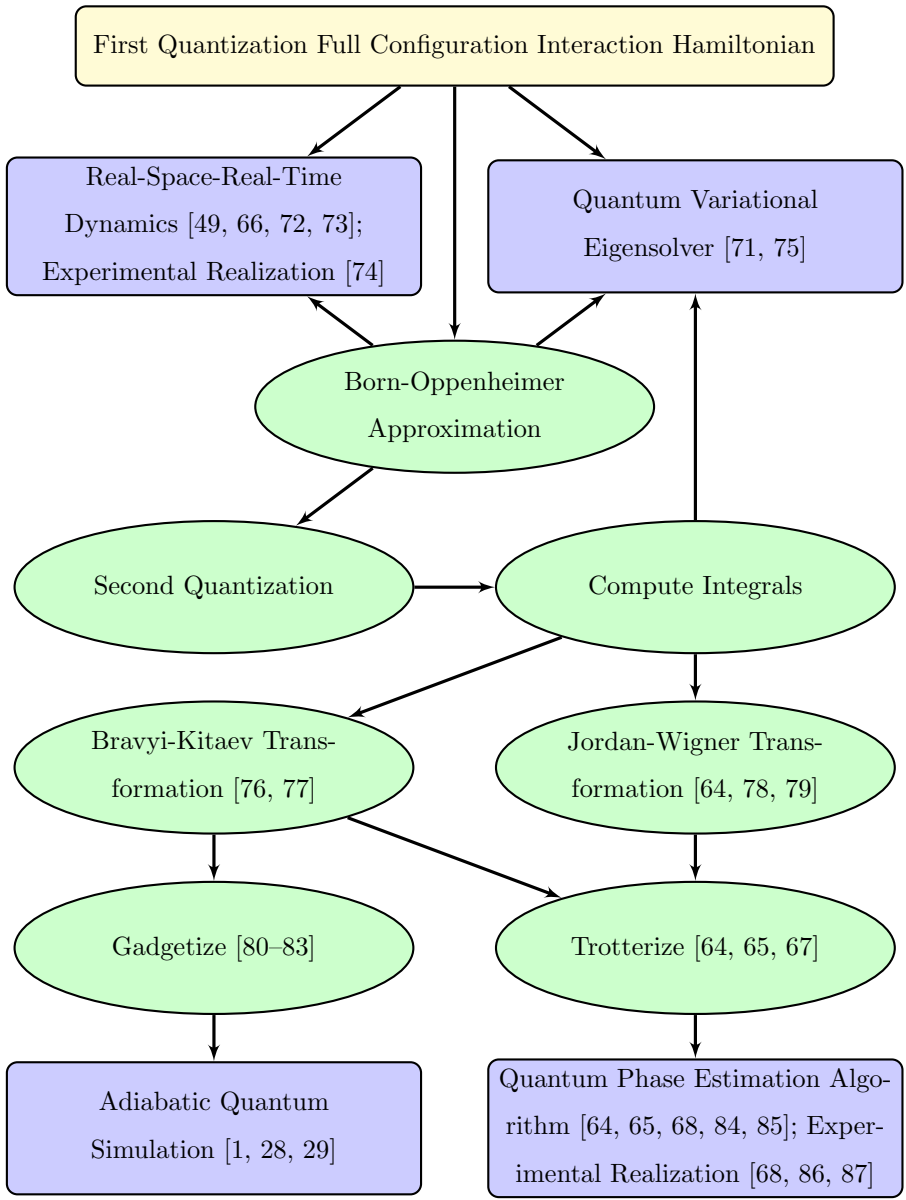
\begin{figure}[H]
\label{flow}
\centering
\begin{tikzpicture}[scale=.8]
\tikzset{top/.style={draw, rectangle, rounded corners, thick, text centered, fill=yellow!20,
text width =28em, minimum height=3.0em}}
\tikzset{block/.style={draw, rectangle, rounded corners, thick, text centered, fill=blue!20,
text width = 15em, minimum height=5em, execute at begin node=\small}}
\tikzset{decision/.style={draw, ellipse, rounded corners, thick, text centered, fill=green!20,
text width = 10em, minimum height=5em, execute at begin node=\small}}
\tikzstyle{line} = [draw, very thick, -latex']
\tikzstyle{vecArrow} = [draw, thick, decoration={markings,mark=at position
   1 with {\arrow[semithick]{open triangle 60}}},
   double distance=1.4pt, shorten >= 5.5pt,
   preaction = {decorate},
   postaction = {draw,line width=1.4pt, white,shorten >= 4.5pt}]

    \node [top] (FCI) {First Quantization Full Configuration Interaction Hamiltonian};

    \node  [block] at (-4,-3) (Spatial) {Real-Space-Real-Time\\ Dynamics  \cite{Kassal2010,Welch2013,Ward2009,Whitfield2013b};\\
Experimental Realization \cite{lu2011}};
    \node [block] at (4,-3) (Var) {Quantum Variational Eigensolver \cite{McClean2013,Yung2013}};

    \node [decision] at (0,-6) (BO) {Born-Oppenheimer Approximation};

    \node [decision] at (-4, -9) (Quant){Second Quantization};
    \node [decision] at (4, -9) (int){Compute Integrals};

    \node [decision] at (-4, -12) (BK){Bravyi-Kitaev Transformation \cite{Seeley2012, Bravyi2000}};
    \node [decision] at (4,-12) (JW){Jordan-Wigner Transformation \cite{Jordan1928,Somma2002,Aspuru-Guzik2006}};

    \node [decision] at (-4, -15) (Gad){Gadgetize \cite{Kempe2004,Jordan2008,Oliveira2005,Cao2013}};
    \node [decision] at (4, -15) (Trot){Trotterize \cite{Aspuru-Guzik2006,Whitfield2010,CodyJones2012}};

    \node [block] at (-4, -18) (AQC) {Adiabatic Quantum Simulation \cite{Farhi2000,Johnson2011,Berkley2013}};
    \node [block] at (4, -18) (PEA) {Quantum Phase Estimation Algorithm \cite{Aspuru-Guzik2006,Lanyon2009,Whitfield2010,Wang2008,Veis2010}; Experimental Realization \cite{Lanyon2009,Li2011,Du2010}};

    \path [line] (FCI) -- (BO);
    \path [line] (FCI) -- (Spatial);
    \path [line] (FCI) -- (Var);

    \path [line] (BO) -- (Var);
    \path [line] (BO) -- (Spatial);
    \path [line] (BO) -- (Quant);

    \path [line] (Quant) -- (int);

    \path [line] (int)  -- (Var) ;
    \path [line] (int) -- (BK);
    \path [line] (int) -- (JW);

    \path [line] (BK) -- (Gad);
    \path [line] (BK) -- (Trot);
    \path [line] (JW) -- (Trot);

    \path [line] (Gad) -- (AQC);
    \path [line] (Trot) -- (PEA);

 \end{tikzpicture}
\caption{\label{hitchhiker}A diagram relating several different approaches to the quantum simulation of quantum chemistry with the procedures and approximations implicit in each approach. Some of these approaches have been demonstrated experimentally using quantum information processors.}
\end{figure}

Lattice systems are a natural target for trapped ion and atom quantum simulators, with the trapping mechanism taking the place of the crystal lattice and interactions restricted to neighbors on the lattice. However, quantum chemistry applied to molecular systems is perhaps the broadest class of problems on which quantum simulation of interacting fermions could have an impact. Finding the energy of electrons interacting in the Coulomb potential of a set of fixed nuclei of an atom or molecule defines the electronic structure problem. This problem appears to be hard for classical computers because the cost of directly solving for the eigenvalues of the exact electronic Hamiltonian grows exponentially with the problem size. In spite of much progress over the last 60 years developing approximate classical algorithms for this problem, exact calculations remain out of reach for many systems of interest. There are many proposals for the efficient quantum simulation of chemical Hamiltonians~\cite{Aspuru-Guzik2006,Whitfield2010,Kassal2010,CodyJones2012,Lanyon2009,Wecker2013,Toloui2013} (see Figure~\ref{hitchhiker}).

One may divide quantum simulation algorithms into two classes: those that address statics and compute ground state properties, and those that address dynamics, and simulate time evolution of the wavefunction. It is clear that the simulation of time evolution is exponentially more efficient on quantum computers, with significant implications for the simulation of chemically reactive scattering, in particular \cite{Kassal2010}. The computation of group state properties naturally requires preparation of the ground state. This can be done adiabatically \cite{Aspuru-Guzik2006}, or by preparation of an ansatz for the ground state~\cite{McClean2013}. Adiabatic preparation of the ground state within a gate model simulation requires time evolution of the wavefunction, which is efficient. However, the length of time for which one must evolve is determined, as for all adiabatic algorithms, by the minimum energy gap between ground and first excited states along the adiabatic path. This is unknown in general. Similarly, a successful ansatz state must have significant overlap with the true ground state, and guarantees of this are unavailable in general. 

The worst case complexity of generic model chemistries (e.g. local fermionic problems studied with density functional theory) has been shown to be in the quantum mechanical equivalent of NP-Complete, QMA-Complete \cite{Schuch2009,Whitfield2013}. However, the subset of these generic models which correspond to stable molecules, or to unstable configurations of chemical interest such as transition states, is small and structured. Just as with adiabatic optimization, it does not matter if molecular electronic structure is QMA-Complete so long as the average instance can be solved (or even approximated) efficiently. In this case we also have considerable heuristic evidence that molecules are able to find their ground state configurations rapidly: these are the configurations in which they naturally occur. Similarly, unstable transition states of interest occur in natural processes. Given that simulation of time evolution on a quantum computer is efficient, we conjecture that simulation of the natural processes that give rise to these states will also be practical.

The proofs that Local Hamiltonian (a decision problem capturing the complexity of finding the ground state energy) is QMA-Complete relies on the construction of various specific Hamiltonians that can represent any possible instance of any problem in QMA. In general, these Hamiltonians possess couplings between more than two qubits. Hamiltonians which contain many-body interactions of order $k$ and lower are referred to as $k$-local Hamiltonians; experimentally programmable couplings are 2-local. The original formulation by Kitaev was $\left(\log n\right)$-local, he then reduced this to $5$-local and that result was subsequently reduced to 3-local. To reduce 3-local Hamiltonians to 2-local Hamiltonians ``perturbative gadgets'' were introduced by Kempe \emph{et al.} \cite{Kempe2004}, which can embed a $k$-local Hamiltonian in a subspace of a 2-local Hamiltonian using ancilla qubits. In the past decade, a growing body of work has pushed the development of different gadgets which embed various target Hamiltonians with various tradeoffs in the resources required \cite{Oliveira2005,Jordan2008,Bravyi2008,Biamonte2007,Biamonte2008,Duan2011,Babbush2013b,Cao2013}.

Embedding problems in realizable Hamiltonians requires careful consideration of the availability of experimental resources. One consideration is that many-body qubit interactions cannot be directly realized experimentally. Another factor is the ``control precision'' of the Hamiltonian which is the dynamic range of field values which a device must be able to resolve in order to embed the intended eigenspectrum to a desired accuracy. This resource is especially important for molecular electronic structure Hamiltonians as chemists are typically interested in acquiring chemical accuracy ($0.04$ eV). Control precision is often the limiting factor when a Hamiltonian contains terms with coefficients which vary by several orders of magnitude. Other considerations include the number of qubits available as well as the connectivity and type of qubit couplings.

In this paper, we describe a scalable method which allows for the application of the quantum adiabatic algorithm to a programmable physical system encoding the molecular electronic Hamiltonian. Our method begins with the second quantized representation of molecular electronic structure in which the Hamiltonian is represented with fermionic creation and annihilation operators. The first step in our protocol is to convert the fermionic Hamiltonian to a qubit Hamiltonian using the Bravyi-Kitaev transformation \cite{Bravyi2000,Seeley2012}. We show that using the Bravyi-Kitaev transformation instead of the Jordan-Wigner transformation is necessary for avoiding exponential control precision requirements in an experimental setting. Next, we show a new formulation of perturbative gadgets motivated by \cite{Kempe2004,Cao2013} that allows us to remove all terms involving $YY$ couplings\footnote{Throughout this paper we use $X$, $Y$ and $Z$ to denote the Pauli matrices. Furthermore, these operators are defined to act as identity on unlabeled registers so that the dot product $Y_i Y_j$ is understood to represent the tensor product $Y_i\otimes Y_j$.} in a single gadget application. Finally, we apply the locality gadgets described in \cite{Jordan2008} to produce a 2-local Hamiltonian with only $ZZ$, $XX$ and $ZX$ couplings.

The paper is organized as follows. We review the second quantized formulation of the electronic structure problem in Section~\ref{sec:2q}. Next we give the mapping of this problem to qubits in Section~\ref{sec:q}. We introduce the gadgets that we will use for locality reduction in Section~\ref{sec:g1}. In Section~\ref{sec:h2} we apply our procedure to a simple example: molecular hydrogen in a minimal basis. We close the paper with some discussion and directions for future work.

\section{Second Quantization}\label{sec:2q}

We begin by writing down the full configuration interaction (FCI) Hamiltonian in the occupation number basis. We define spin orbitals as the product of a spin function (representing either spin up or spin down) and a single-electron spatial function (usually molecular orbitals produced from a Hartree-Fock calculation). For example, in the case of molecular hydrogen there are two electrons and thus, two single-electron molecular orbitals, $|\psi_1\rangle$ and $|\psi_2\rangle$. Electrons have two possible spin states, $|\alpha\rangle$ (spin up) and $|\beta\rangle$ (spin down). The four spin orbitals for molecular hydrogen are therefore, $|\chi_0\rangle = |\psi_1\rangle|\alpha\rangle$, $|\chi_1\rangle = |\psi_1\rangle|\beta\rangle$, $|\chi_2\rangle = |\psi_2\rangle|\alpha\rangle$, and $|\chi_3\rangle = |\psi_2\rangle|\beta\rangle$.

The occupation number basis is formed from all possible configurations of $n$ spin orbitals which are each either empty or occupied. We represent these vectors as a tensor product of individual spin orbitals written as $|f_{n-1} ... f_0\rangle$ where $f_j \in \mathbb{B}$ indicates the occupation of spin orbital $|\chi_j\rangle$. Any interaction between electrons can be represented as some combination of creation and annihilation operators $a_j^\dagger$ and $a_j$ for $\{j \in \mathbb{Z}| 0 \leq j < n\}$. Because fermionic wavefunctions must be antisymmetric with respect to particle label exchange, these operators must obey the fermionic anti-commutation relations\footnote{The anti-commutator of operators $A$ and $B$ is defined as, $\left[A,B\right]_+ \equiv A\cdot B + B \cdot A$.},
\begin{equation}
\label{femcom}
[a_j, a_k]_+ = [a_j^\dagger, a_k^\dagger]_+ = 0,
\quad [a_j, a_k^\dagger]_+ = \delta_{jk}\openone.
\end{equation}
With these definitions we write the second-quantized molecular electronic Hamiltonian,
\begin{equation}
\label{second}
H = \sum_{i,j} h_{ij} a_i^\dagger a_j + \frac{1}{2} \sum_{i,j,k,l} h_{ijkl} a_i^\dagger a_j^\dagger a_k a_l.
\end{equation}
The coefficients $h_{ij}$ and $h_{ijkl}$ are single and double electron overlap integrals which are precomputed classically. The number of distinct integrals scale as $O\left(n^4\right)$ in the number of molecular orbitals $n$.

\section{Qubit Representation}\label{sec:q}

The next step in our reduction will be to represent our fermionic wavefunction in terms of qubits. We use the direct mapping introduced in \cite{Aspuru-Guzik2006} that maps an occupancy state to a qubit basis state.  Using Pauli operators we can represent qubit raising and lowering operators as,
\begin{equation}
Q_j^+ = |1\rangle\langle 0| = \frac{1}{2}\left(X_j - i Y_j\right), 
\quad Q_j^- = |0\rangle\langle 1| = \frac{1}{2}\left(X_j + i Y_j\right).
\end{equation}
However, these operators do not obey the fermionic commutation relations given in Eq.~\ref{femcom}. To write qubit operators that obey the commutation relations in Eq.~\ref{femcom}, we could use the Jordan-Wigner transformation \cite{Jordan1928,Somma2002,Aspuru-Guzik2006}.

Unfortunately, the Jordan-Wigner transformation is not a scalable way to reduce electronic structure to an experimentally realizable Hamiltonian for AQC. This is because the Jordan-Wigner transformation introduces $k$-local interaction terms into the Hamiltonian and $k$ grows linearly in the system size. Prima facie, this is not a major problem because there exist theoretical tools known as perturbative gadgets (the focus of Section~\ref{sec:g1}) which allow for reductions in interaction order. However, in all known formulations of perturbative gadgets, control precision increases exponentially in $k$. Thus, the linear locality overhead introduced by the Jordan-Wigner transformation translates into an exponential control precision requirement in the reduction.

An alternative mapping between the occupation number basis and qubit representation, known as the Bravyi-Kitaev transformation, introduces logarithmic locality overhead \cite{Bravyi2000,Seeley2012}. Two pieces of information are required in order to correctly construct creation and annihilation operators that act on qubits but which obey the fermionic commutation relations. Firstly, the occupancy of each orbital must be stored. Secondly, parity information must be stored so that, for a pair of orbitals $i$ and $j$ it is possible to determine the parity of the occupancy of the orbitals that lie between them. This parity determines the phase which results from exchanging the occupancy of the two orbitals.

The occupation number basis stores the occupation directly in the qubit state (hence the name). This implies that occupancy is a fully local variable in this basis; one may determine the occupancy of an orbital by measuring a single qubit. However, this also implies that the parity information is completely non-local. It is this fact that determines the structure of the qubit creation and annihilation operators in the Jordan-Wigner transformation. Each such operator changes the state of a single qubit $j$ (updating the occupancy information) but also acts on all qubits with indices less than $j$ to determine the parity of their occupancy. This results in qubit operators, expressed as tensor products of Pauli matrices, that contain strings of $Z$ operators whose length grows with the number of qubits. One could consider storing the parity information locally, so that the qubit basis states store sums of orbital occupancies. Then determination of parity requires a single qubit operation. However, updating occupancy information requires updating the state of a number of qubits that again grows with the number of qubits. Hence this ``parity basis'' construction offers no advantage over the Jordan Wigner transformation \cite{Bravyi2000}.

The Bravyi-Kitaev transformation offers a middle ground in which both parity and occupancy information are stored non-locally, so neither can be determined by measurement of a single qubit \cite{Bravyi2000,Seeley2012}. Both parity and occupancy information can be accessed by acting on a number of qubits that scales as the logarithm of the number of qubits. It is this logarithmic scaling that makes the proposed mapping of electronic structure to a 2-local qubit Hamiltonian efficient.

The consequences of this mapping, originally defined in \cite{Bravyi2000}, were computed for electronic structure in \cite{Seeley2012}. That work defines several subsets of qubits in which the parity and occupancy information is stored. The occupancy information is stored in the update set, whereas the parity information is stored in the parity set. These sets are distinct and their size is strictly bounded above by the logarithm base two of the number of qubits. The total number of qubits on which a qubit creation and annihilation operator may act can be a multiple of the logarithm base two of the number of qubits. However, this multiple is irrelevant from the point of view of the scalability of the construction. Using the Bravyi-Kitaev transformation, the spin Hamiltonian for molecular hydrogen in the minimal (STO-3G) basis, as reported in \cite{Seeley2012}, is given by
\begin{eqnarray}
\label{H2}
H_{\textrm{H}_2} & = & f_0 \openone + f_1 Z_0 + f_2 Z_1 + f_3Z_2 +  f_1 Z_0   Z_1 + f_4 Z_0   Z_2 +  f_5 Z_1   Z_3\\
& + &  f_6 X_0   Z_1   X_2  + f_6Y_0  Z_1   Y_2 + f_7 Z_0   Z_1   Z_2 + f_4 Z_0   Z_2   Z_3 + f_3 Z_1   Z_2   Z_3\nonumber\\
& + & f_6 X_0   Z_1   X_2   Z_3 + f_6 Y_0   Z_1   Y_2   Z_3 + f_7 Z_0   Z_1  Z_2  Z_3\nonumber
\end{eqnarray}
and
\begin{eqnarray}
& f_0 = -0.812610, \quad f_1 = 0.171201, \quad f_2 = 0.168623, \quad f_3 = -0.222780 &\\
& f_4 = 0.120546, \quad f_5 = 0.174349, \quad f_6 = 0.0453218, \quad f_7 = 0.165868.& \nonumber
\end{eqnarray}
In general, the Bravyi-Kitaev transformation applied to electronic structure produces an $n$-qubit Hamiltonian which is $(\log n)$-local, and has $n^4$ real terms. This implies that each term has an even number of $Y$ terms, or none.

\section{Hamiltonian Gadgets}\label{sec:g1}

In order to embed electronic structure in an experimentally realizable Hamiltonian, we define a scalable methodology for transforming our $\left(\log n\right)$-local qubit Hamiltonian into a 2-local Hamiltonian with only $ZZ$, $XX$ and $XZ$ interaction terms.  In this section we will describe tools known as ``gadgets'' which allow us to simulate the target Hamiltonian with these interactions. Specifically, gadgets provide a method for embedding the eigenspectra (and sometimes eigenvectors) of an $n$-qubit ``target'' Hamiltonian, denoted by $H_\textrm{target}$, in a restricted (typically low-energy) subspace of a more constrained $\left(N > n\right)$-qubit ``gadget'' Hamiltonian, denoted by $\widetilde{H}$.

To illustrate the general idea of gadgets, we describe how a $2$-local Hamiltonian can embed a $k$-local Hamiltonian. Suppose that we have a gadget Hamiltonian, $\widetilde{H}$, which contains only 2-local terms which act on $N = n + a$ qubits. Then,
\begin{equation}
\widetilde{H} = \sum_{i = 1} f_{i} O_i, \quad \widetilde{H} |\psi_i\rangle = \widetilde{\lambda}_i |\widetilde{\psi}_i\rangle,
\end{equation}
where $\{f_i\}$ are scalar coefficients, $\widetilde{\lambda}_j$ and $|\widetilde{\psi}_i\rangle$ are the eigenvectors and eigenvalues of $\widetilde{H}$, and $\{O_i\}$ are the 2-local interaction terms of the physical Hamiltonian. We choose our interaction terms to be Hilbert-Schmidt orthogonal so that $\textrm{Tr}\left[O_i O_j\right] = 2^n \delta_{i,j}$. We now define an \emph{effective Hamiltonian} which has support on the lowest $2^n$ states of the gadget,
\begin{equation}
H_\textrm{eff}  \equiv \sum_{i=0}^{2^n - 1} \widetilde{\lambda}_i |\widetilde{\psi}_i\rangle\langle \widetilde{\psi}_i |  =  \sum_{i = 1} f_i O_i \otimes \Pi.\end{equation}
Here $\Pi$ is a projector onto a particular state (usually the lowest energy state) of the $a$ ancilla qubits and the $\{O_i\}$ are a Hilbert-Schmidt orthogonal operator basis for operators on the space of the $n$ logical qubits. In other words, the most general representation of $H_\textrm{eff}$ is an expansion of all possible tensor products acting on the logical qubits. In general, there is no reason why $f_i = 0$ on all non-2-local terms. Therefore a 2-local gadget on $N = n + a$ qubits can embed a $\left(k > 2\right)$-local, $n$-qubit Hamiltonian using $a$ ancilla bits. Determination of the form of the effective Hamiltonian requires techniques that do not rely on construction of the full effective Hamiltonian for a many qubit system, and so perturbative techniques have been used extensively.
\begin{figure}[H]
\begin{subfigure}[b]{.5\textwidth}
\resizebox{\linewidth}{!}{
\includegraphics{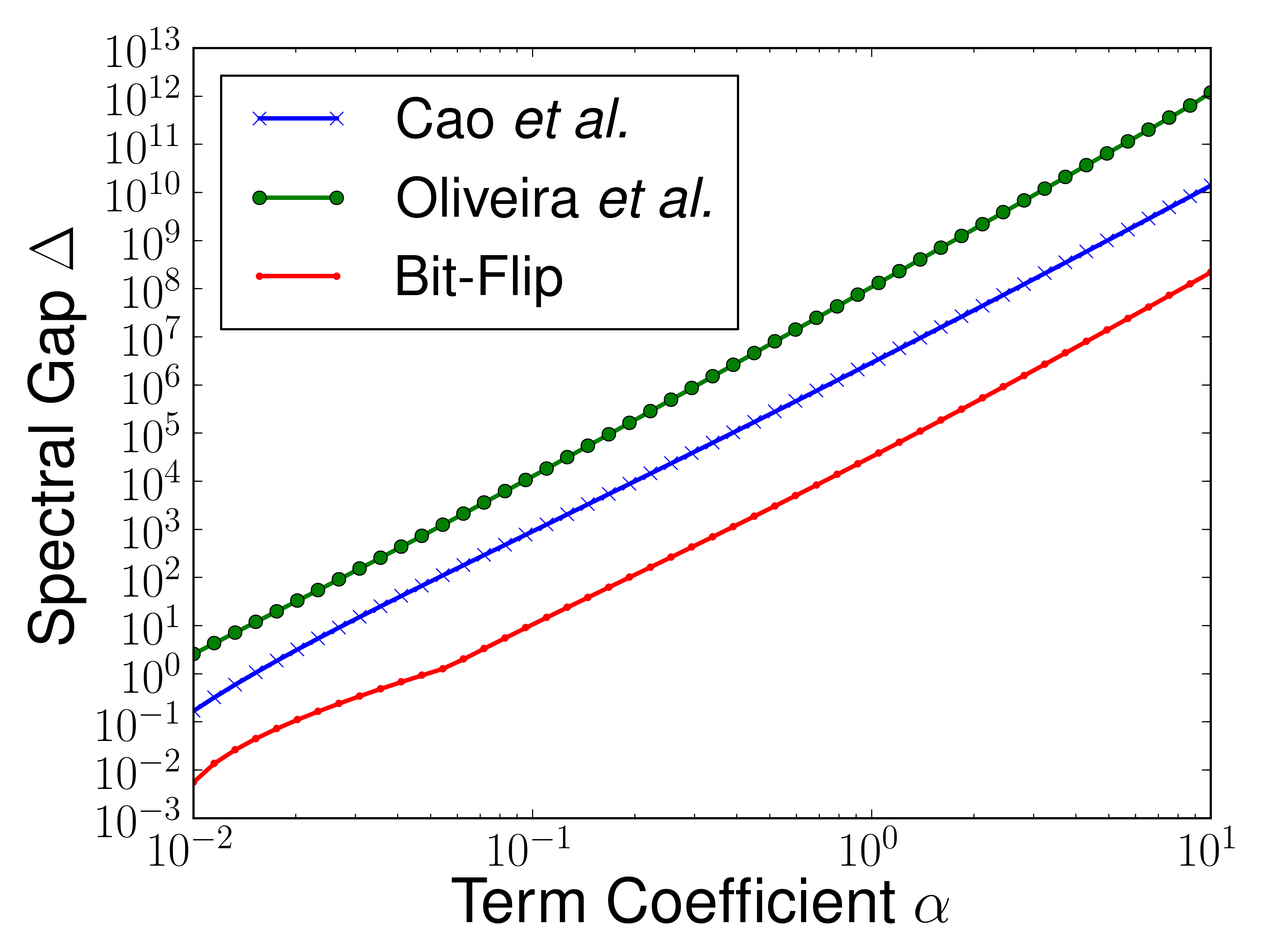}}
\end{subfigure}
\begin{subfigure}[b]{.5\textwidth}
\resizebox{\linewidth}{!}{
\includegraphics{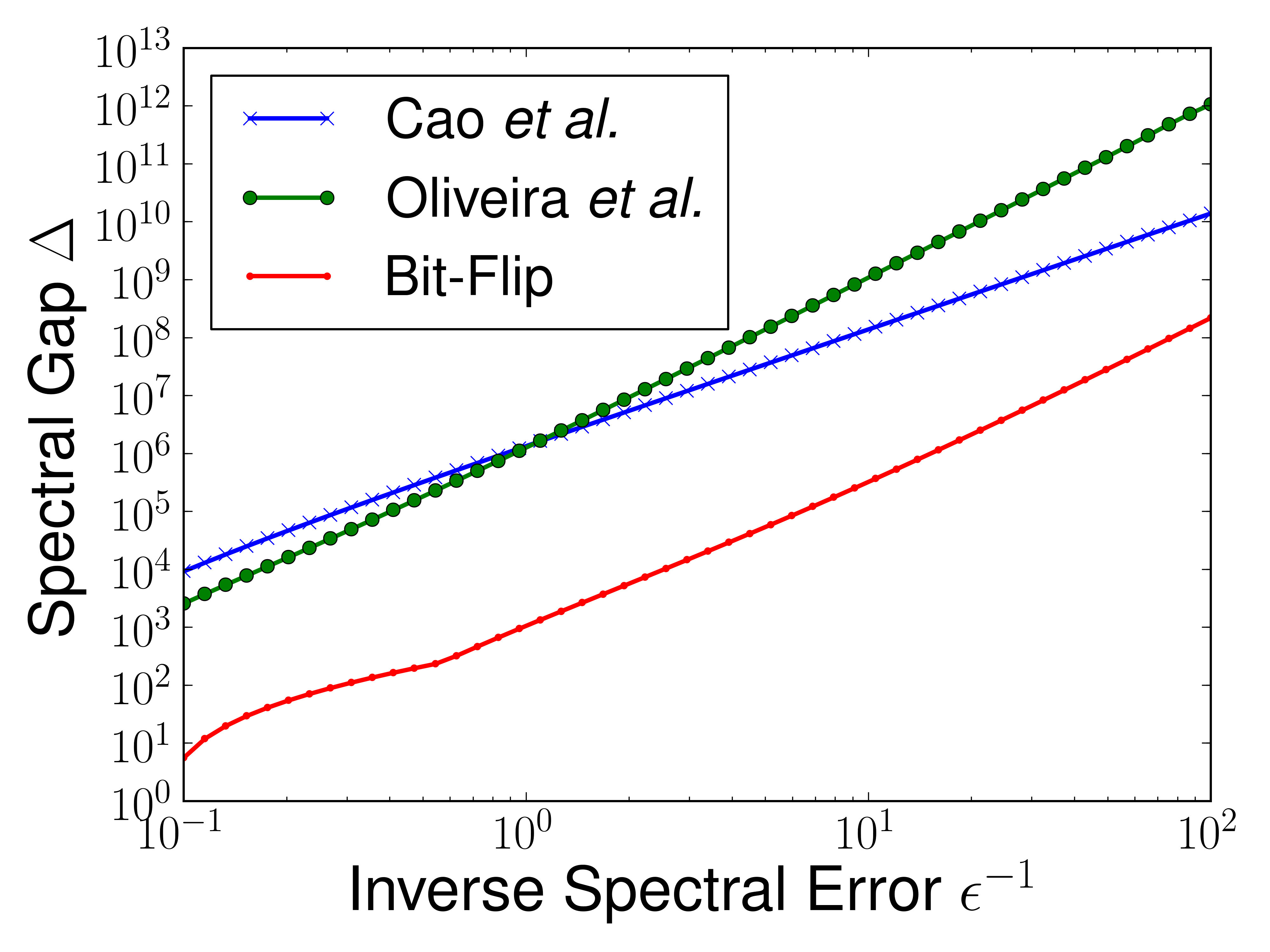}}
\end{subfigure}
\caption{\label{gadcompare} Numerics comparing the minimum spectral gaps required to reduce the term $\alpha X_1Y_2 Z_3$ to 2-local with an error in the eigenspectrum of at most $\epsilon$. On the left, $\epsilon$ is fixed at 0.01 Hartree and gaps are plotted as a function of $\alpha$. On the right, $\alpha$ is fixed at 10 Hartree and gaps are plotted as a function of $\epsilon^{-1}$. Here we compare the bit-flip construction \cite{Kempe2004,Jordan2008}, the Oliveira and Terhal construction \cite{Oliveira2005} and an improved variant on Oliveira and Terhal by Cao \emph{et al.} \cite{Cao2013}. Further comparisons of this nature are made in \cite{Cao2013}.}
\end{figure}

The use of perturbation theory to derive Hamiltonian gadgets was introduced by Kempe \emph{et al.} in their canonical proof showing that 2-Local Hamiltonian is QMA-Complete \cite{Kempe2004}. Their construction, which we refer to as the ``bit-flip construction'' for reasons that will become obvious later on, was analyzed by Jordan and Farhi using a formulation of perturbation theory due to Bloch \cite{Jordan2008}. Other perturbative gadget constructions were introduced by Oliveira and Terhal to prove the QMA-Completeness of Hamiltonian on a square lattice \cite{Oliveira2005}. Following this work, Biamonte and Love used gadgets to show that $XX$ and $ZZ$, or $XZ$ couplings alone, suffice for the QMA-Completeness of 2-local Hamiltonian \cite{Biamonte2007}. Several other papers improve these gadgets from an experimental perspective and introduce novel constructions which are compatible with the protocol developed here~\cite{Biamonte2008,Bravyi2008,Duan2011,Babbush2013b,Cao2013}. We note that different types of gadgets may have specific advantages when designing Hamiltonians for specific hardware. Results from \cite{Cao2013} suggest that there is a rough tradeoff between the number of ancillae required and the amount of control precision required. For instance, Figure~\ref{gadcompare} indicates that bit-flip gadgets require less control precision than other gadget constructions (but generally more ancillae). In this paper we focus on the bit-flip family of gadgets.

Although we employ the perturbation theory approach here, it does require a high degree of control precision and should be avoided when possible. We point out that when the Hamiltonian is entirely diagonal there are exact gadgets \cite{Biamonte2008,Babbush2013b} which can embed the ground state with far less control precision and often far fewer ancillae but in a way that does not necessarily conserve the gap scaling. Moreover, ``frustration-free'' gadgets have been used extensively in proofs of the QMA-Completeness of various forms of quantum satisfiability, and in restricting the necessary Hamiltonian terms for universal adiabatic quantum computing~\cite{Nagaj2007,Nagaj2010,Gosset2013,Childs2013}.

While several types of perturbation theory have been used to derive these gadgets, we closely follow the approach and notation of Kempe \emph{et al.} \cite{Kempe2004}. We wish to analyze the spectrum of the gadget Hamiltonian, $\widetilde{H} = H + V$ for the case that the norm of the perturbation Hamiltonian, $V$, is small compared to the spectral gap between the ground state and first excited state of the unperturbed Hamiltonian, $H$. To accomplish this we use the Green's function of $\widetilde{H}$,
\begin{equation}
\label{greens}
\widetilde{G}\left(z\right) \equiv \left(z \openone - \widetilde{H}\right)^{-1} = \sum_{j}\frac{ |\widetilde{\psi}_j\rangle\langle\widetilde{\psi}_j|}{z - \widetilde{\lambda}_j}.
\end{equation}
We also define $G\left(z\right)$ using the same expression except with $H$ instead of $\widetilde{H}$. Further, let ${\cal H} = {\cal L}_+ \oplus {\cal L}_-$ be the Hilbert space of $\widetilde{H}$ where ${\cal L}_+$ is the ``high-energy'' subspace spanned by eigenvectors of $\widetilde{H}$ with eigenvalues $\widetilde{\lambda} \geq \lambda_*$ and ${\cal L}_-$ is the complementary ``low-energy'' subspace, spanned by eigenvectors of $\widetilde{H}$ corresponding to eigenvalues of $\widetilde{\lambda} < \lambda_*$. Let $\Pi_\pm$ correspond to projectors onto the support of ${\cal L_\pm}$. In a representation of ${\cal H} =  {\cal L}_+ \oplus {\cal L}_-$, all the aforementioned operators $V$, $H$, $\widetilde{H}$, $G\left(z\right), \widetilde{G}\left(z\right)$ are block-diagonal so we employ the notation that $A_{\pm \pm} = \Pi_\pm A \, \Pi_\pm$ and,
\begin{equation}\label{Aeqn}
A = \left(\begin{matrix} A_+ & A_{+-} \\ A_{-+} & A_- \\ \end{matrix}\right).
\end{equation}
Finally, we define the operator-valued function known as the \emph{self-energy},
\begin{equation}
\label{Sigma}
\Sigma_-\left(z\right) \equiv z \openone_- - \widetilde{G}_-^{-1}\left(z\right).
\end{equation}
We now use this notation to restate the ``gadget theorem'' proved as \emph{Theorem 6.2} in \cite{Kempe2004}.
\newtheorem{theorem}{Theorem}
\begin{theorem}
[Theorem 6.2 in \cite{Kempe2004}] Assume that $H$ has a spectral gap $\Delta$ around the cutoff $\lambda_*$; i.e. all of its eigenvalues are in $\left(-\infty, \lambda_-\right] \cup \left[\lambda_+, +\infty\right)$ where $\lambda_+= \lambda_* + \Delta/2$ and $\lambda_- = \lambda_* - \Delta/2$. Assume that $\| V \| \leq \Delta/2$. Let $\epsilon > 0$ be arbitrary. Assume there exists an operator $H_\textrm{eff}$ such that $\lambda\left(H_\textrm{eff}\right) \subset \left[c,d\right]$ for some $ c < d < \lambda_*-\epsilon$ and, moreover, the inequality $\| \Sigma_-\left(z\right) - H_\textrm{eff}\| \leq \epsilon$ holds for all $z \in \left[c-\epsilon, d + \epsilon\right]$. Then each eigenvalue $\widetilde{\lambda}_j$ of $\widetilde{H}_-$ is $\epsilon$-close to the $j^\textrm{th}$ eigenvalue of $H_\textrm{eff}$.
\end{theorem}

Theorem 1 assures us that the eigenspectrum of the self-energy provides an arbitrarily good approximation to the eigenspectrum of the low-energy subspace of the gadget Hamiltonian. This is useful because the self-energy admits a series expansion,
\begin{equation}
\Sigma_-\left(z\right) = H_- + V_- + \sum_{k = 2}^{\infty} V_{-+} G_+\left(V_+ G_+\right)^{k-2} V_{+-}.
\end{equation}
Using $G_+ = \left(z - \Delta\right)^{-1} \openone_+ $ and $H_-=0$, we focus on the range $z = O\left(1\right) \ll \Delta$ and find that,
\begin{equation}
\label{eff}
H_\textrm{eff} \approx V_- + \frac{1}{\Delta}\sum_{k = 2}^{\infty} V_{-+} \left(\frac{V_+}{\Delta}\right)^{k-2}\!\!\! V_{+-}.
\end{equation}

We use this effective Hamiltonian to approximate our $k$-local target Hamiltonian, which we now specify. The terms in our target Hamiltonian will have a locality that scales logarithmically with the number of orbitals. We may write such a term:
\begin{equation}
T=\bigotimes_{i=0}^{k-1} O_i ~ : ~ O_i \in\{X_i,Y_i,Z_i\}~\forall i.
\end{equation}
One can always apply gadgets term by term to reduce locality; however, this may not be the optimal procedure. In addition, we are interested in replacing even tensor powers of the $Y$ operator. For both these reasons we consider a slightly more general form of term as a target for gadgetization. We use the fact that it is only the commuting nature of the $\{O_i\}$ that is important for the gadget to function. We therefore write our target term as a product of $k$ commuting operators, which includes the special case in which it is a product of $k$ operators acting on distinct tensor factors,
\begin{equation}
T' =\prod_{i=0}^{k-1} O_i ~ : ~ \left[O_i, O_j\right] = 0~\forall \{i,j\}
\end{equation}
Hence, we can represent the target Hamiltonian as a sum of $r$ terms which are the product of $k$ commuting operators,
\begin{equation}
H_\textrm{target} = H_\textrm{else} + \sum_{s = 1}^{r} \prod_{i = 0}^{k-1} O_{s,i}
\end{equation} 
where all $\{O_{s,i}\}$ commute for a given $s$ and $H_\textrm{else}$ can be realized directly by the physical Hamiltonian. While previous formulations of bit-flip gadgets \cite{Kempe2004,Jordan2008,Cao2013} have gadgetized operators acting on distinct tensor factors, it is only necessary that the operators commute. Their action on distinct tensor factors is sufficient but not necessary for the gadget construction. We take advantage of this property in order to realize $YY$ terms without access to such couplings by making the substitution, $Y_i Y_j  \rightarrow - X_i X_j Z_i Z_j$. Since $X_i X_j$ commutes with $Z_i Z_j$, we can create this effective interaction with a bit-flip gadget. For instance, suppose we have the term, $Z_0 Y_1 Y_2$. We gadgetize the term $A \cdot B \cdot C$ where $A = Z_0$, $B = -X_1 X_2$, and $C = Z_1 Z_2$ and all operators $A,B,C$ commute. We note that another approach to removing $YY$ terms is explained in \cite{Cao2013}.

We now introduce the form of the penalty Hamiltonian that acts only on the ancilla qubits. Bit-flip gadgets introduce an ancilla system which has two degenerate ground-states, usually taken to be $|111...\rangle_u$ and $|000...\rangle_u$ where $u$ indicates that these kets refer to an ancilla space. For each of the $r$ terms we use a separate ancilla system of the form,
\begin{equation}
H_s = \frac{\Delta}{2\left(k-1\right)}\sum_{0 \leq i < j \leq k-1} \!\! \left(\openone - Z_{u_{s,i}} Z_{u_{s,j}} \right).
\end{equation}
Again, we use $u$ to indicate that operators act on an ancilla; e.g. the label $u_{3,2}$ indicates the ancilla corresponding to $O_{3,2}$ (the second operator in the third term). For each term we follow Farhi and Jordan in introducing an ancilla system connected by a complete graph with equal and negative edge weights. Thus, the ground state of the ancilla system is spanned by $|111...\rangle_u$ and $|000...\rangle_u$. 

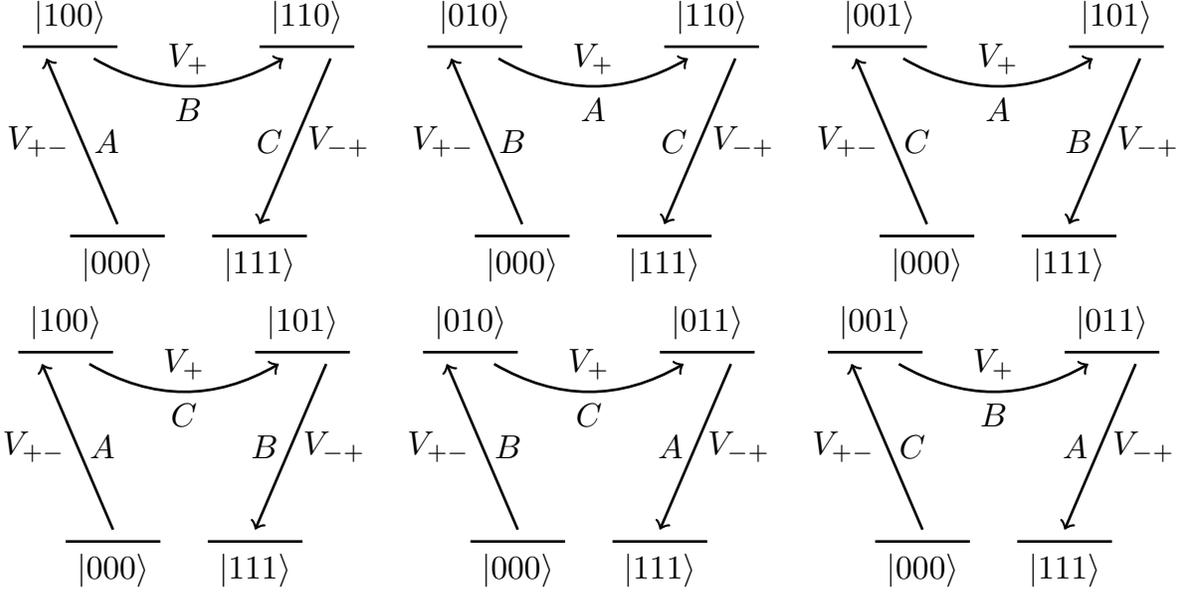
\begin{figure}[h]
\centering
\begin{subfigure}[b]{0.32\textwidth}
\resizebox{\linewidth}{!}{
    \begin{tikzpicture}[
      scale=0.5,
      level/.style={thick},
      trans/.style={->, thick},
      plus/.style={->, bend right, thick},
    ]
    \draw[level] (1,0) -- (3,0) node[midway, below] {$\ket{000}$};
    \draw[level] (0,4) -- (2,4) node[midway, above] {$\ket{100}$};
    \draw[level] (5,4) -- (7,4) node[midway, above] {$\ket{110}$};
    \draw[level] (4,0) -- (6,0) node[midway, below] {$\ket{111}$};
    \draw[trans] (2,0.25) to node[left] {$V_{+-}$} node[right] {$A$} (0.5,3.75);
    \draw[plus] (1.5,3.75) to node[above] {$V_{+}$} node[below] {$B$} (5.5,3.75);
    \draw[trans] (6.5,3.75) to node[right] {$V_{-+}$}  node[left] {$C$} (5,0.25);
    \end{tikzpicture}}
\end{subfigure}
\begin{subfigure}[b]{0.32\textwidth}
\resizebox{\linewidth}{!}{
    \begin{tikzpicture}[
      scale=0.5,
      level/.style={thick},
      trans/.style={->, thick},
      plus/.style={->, bend right, thick},
    ]
    \draw[level] (1,0) -- (3,0) node[midway, below] {$\ket{000}$};
    \draw[level] (0,4) -- (2,4) node[midway, above] {$\ket{010}$};
    \draw[level] (5,4) -- (7,4) node[midway, above] {$\ket{110}$};
    \draw[level] (4,0) -- (6,0) node[midway, below] {$\ket{111}$};
    \draw[trans] (2,0.25) to node[left] {$V_{+-}$} node[right] {$B$} (0.5,3.75);
    \draw[plus] (1.5,3.75) to node[above] {$V_{+}$} node[below] {$A$} (5.5,3.75);
    \draw[trans] (6.5,3.75) to node[right] {$V_{-+}$}  node[left] {$C$} (5,0.25);
    \end{tikzpicture}}
\end{subfigure}
\begin{subfigure}[b]{0.32\textwidth}
\resizebox{\linewidth}{!}{
    \begin{tikzpicture}[
      scale=0.5,
      level/.style={thick},
      trans/.style={->, thick},
      plus/.style={->, bend right, thick},
    ]
    \draw[level] (1,0) -- (3,0) node[midway, below] {$\ket{000}$};
    \draw[level] (0,4) -- (2,4) node[midway, above] {$\ket{001}$};
    \draw[level] (5,4) -- (7,4) node[midway, above] {$\ket{101}$};
    \draw[level] (4,0) -- (6,0) node[midway, below] {$\ket{111}$};
    \draw[trans] (2,0.25) to node[left] {$V_{+-}$} node[right] {$C$} (0.5,3.75);
    \draw[plus] (1.5,3.75) to node[above] {$V_{+}$} node[below] {$A$} (5.5,3.75);
    \draw[trans] (6.5,3.75) to node[right] {$V_{-+}$}  node[left] {$B$} (5,0.25);
    \end{tikzpicture}}
\end{subfigure}
\begin{subfigure}[b]{0.32\textwidth}
\resizebox{\linewidth}{!}{
    \begin{tikzpicture}[
      scale=0.5,
      level/.style={thick},
      trans/.style={->, thick},
      plus/.style={->, bend right, thick},
    ]
    \draw[level] (1,0) -- (3,0) node[midway, below] {$\ket{000}$};
    \draw[level] (0,4) -- (2,4) node[midway, above] {$\ket{100}$};
    \draw[level] (5,4) -- (7,4) node[midway, above] {$\ket{101}$};
    \draw[level] (4,0) -- (6,0) node[midway, below] {$\ket{111}$};
    \draw[trans] (2,0.25) to node[left] {$V_{+-}$} node[right] {$A$} (0.5,3.75);
    \draw[plus] (1.5,3.75) to node[above] {$V_{+}$} node[below] {$C$} (5.5,3.75);
    \draw[trans] (6.5,3.75) to node[right] {$V_{-+}$}  node[left] {$B$} (5,0.25);
    \end{tikzpicture}}
\end{subfigure}
\begin{subfigure}[b]{0.32\textwidth}
\resizebox{\linewidth}{!}{
    \begin{tikzpicture}[
      scale=0.5,
      level/.style={thick},
      trans/.style={->, thick},
      plus/.style={->, bend right, thick},
    ]
    \draw[level] (1,0) -- (3,0) node[midway, below] {$\ket{000}$};
    \draw[level] (0,4) -- (2,4) node[midway, above] {$\ket{010}$};
    \draw[level] (5,4) -- (7,4) node[midway, above] {$\ket{011}$};
    \draw[level] (4,0) -- (6,0) node[midway, below] {$\ket{111}$};
    \draw[trans] (2,0.25) to node[left] {$V_{+-}$} node[right] {$B$} (0.5,3.75);
    \draw[plus] (1.5,3.75) to node[above] {$V_{+}$} node[below] {$C$} (5.5,3.75);
    \draw[trans] (6.5,3.75) to node[right] {$V_{-+}$}  node[left] {$A$} (5,0.25);
    \end{tikzpicture}}
\end{subfigure}
\begin{subfigure}[b]{0.32\textwidth}
\resizebox{\linewidth}{!}{
    \begin{tikzpicture}[
      scale=0.5,
      level/.style={thick},
      trans/.style={->, thick},
      plus/.style={->, bend right, thick},
    ]
    \draw[level] (1,0) -- (3,0) node[midway, below] {$\ket{000}$};
    \draw[level] (0,4) -- (2,4) node[midway, above] {$\ket{001}$};
    \draw[level] (5,4) -- (7,4) node[midway, above] {$\ket{011}$};
    \draw[level] (4,0) -- (6,0) node[midway, below] {$\ket{111}$};
    \draw[trans] (2,0.25) to node[left] {$V_{+-}$} node[right] {$C$} (0.5,3.75);
    \draw[plus] (1.5,3.75) to node[above] {$V_{+}$} node[below] {$B$} (5.5,3.75);
    \draw[trans] (6.5,3.75) to node[right] {$V_{-+}$}  node[left] {$A$} (5,0.25);
    \end{tikzpicture}}
\end{subfigure}
    \vspace{-.25cm}
  \caption{\label{sixfig}The six equivalent bit-flip processes at third order which produce the effective interaction $A\cdot B \cdot C$. Each of these diagrams also occurs backwards on the part of the ground state in $\ket{111}$.}
\end{figure}

Next, we introduce the perturbation Hamiltonian,
\begin{equation}
V = H_\textrm{else} + \Lambda + \mu_k \sum_{s = 1}^r \sum_{i = 0}^{k - 1} O_{s,i}  X_{u_{s,i}},
\end{equation}
where  $\mu_k = \sqrt[k]{\frac{\Delta^{k-1}}{k!}}$ and $\Lambda$ is a 2-local operator on logical bits which will be discussed later. The effect of this Hamiltonian on the low energy subspace is to introduce virtual excitations into the high energy space that modify the low energy effective Hamiltonian. Only terms which start and end in the ground state contribute to the perturbation series for the self-energy (see, for example, Figure~\ref{sixfig}). Thus, the gadget will produce the target term at order $k$ in which a transition between the two degenerate ground states of the ancillae requires that each of the $X_u$ terms in the perturbation act exactly once to flip all $r\cdot k$ bits from one ground state to the other. Crucially, the order in which the ancillae are flipped does not matter since the operators $O_{s,i}$ commute for a given $s$. The complete gadget is
\begin{equation}
\widetilde{H} = \Lambda + H_\textrm{else} + \sum_{s = 1}^{r} \left(
\frac{\Delta}{2\left(k-1\right)}\sum_{0 \leq i < j \leq k-1}\!\! \left(\openone - Z_{u_{s,i}} Z_{u_{s,j}} \right)
+ \left(\sqrt[k]{\frac{\Delta^{k-1}}{k!}}\right)\sum_{i = 0}^{k - 1} O_{s,i} X_{u_{s,i}}\right)
\end{equation}
and is related to the target Hamiltonian and effective Hamiltonian by,
\begin{equation}
\widetilde{H}_- = H_\textrm{target} \otimes \Pi_- =  H_\textrm{eff}
\end{equation}
for the appropriate choice of $\Lambda$ and $\Delta \gg \| V \|$ where $\Pi_-$ projects onto the ancillae ground space,
\begin{equation}
\Pi_- = \ket{000}\bra{000}_u + \ket{111}\bra{111}_u.
\end{equation}
To illustrate the application of such a gadget and demonstrate how $\Lambda$ is chosen, we scalably reduce the locality of molecular hydrogen and remove all $Y$ terms in Section~\ref{sec:h2}.

As an example, consider the target, $H_\textrm{target} = A\cdot B \cdot C + H_\textrm{else}$. The perturbation is given by,
\begin{equation}
V = \mu A X_a+ \mu B X_b + \mu C X_c + H_\textrm{else} + \Lambda.
\end{equation}
Its components in the low energy subspace, as in the block diagonal representation of Eq.~\ref{Aeqn} is:
\begin{equation}
V_- = \left(H_\textrm{else} + \Lambda\right)\otimes\left(|000\rangle\langle 000|_u + |111\rangle \langle 111|_u\right).
\end{equation}
The projection into the high energy subspace is:
\begin{eqnarray}
V_+ & = & \left(H_\textrm{else} + \Lambda\right)\otimes\left(\sum_{\{a,b,c\} \in \mathbb{B}^3} |a,b,c\rangle \langle a,b,c|_u\right) - V_-\\
& + & \mu A \otimes \left(|0,1,0\rangle\langle 1,1,0|_u+|1,1,0\rangle\langle|0,1,0|_u + |0,0,1\rangle\langle 1,0,1|_u + |1,0,1\rangle\langle 0,0,1|_u \right)\nonumber\\
& + & \mu B \otimes \left(|1,0,0\rangle\langle 1,1,0|_u+|1,1,0\rangle\langle|1,0,0|_u + |0,0,1\rangle\langle 0,1,1|_u + |0,1,1\rangle\langle 0,0,1|_u \right)\nonumber\\
& + & \mu C \otimes \left(|1,0,0\rangle\langle 1,0,1|_u+|1,0,1\rangle\langle|1,0,0|_u + |0,1,0\rangle\langle 0,1,1|_u + |0,1,1\rangle\langle 0,1,0|_u \right)\nonumber.
\end{eqnarray}
The projections coupling the low energy and high energy subspace are:
\begin{eqnarray}
V_{+-} & = & \mu A\otimes\left(|1,0,0\rangle\langle 0,0,0|_u + |0,1,1\rangle\langle 1,1,1|_u\right)\\
& + & \mu B\otimes\left(|0,1,0\rangle\langle 0,0,0|_u + |1,0,1\rangle\langle 1,1,1|_u\right)\nonumber\\
& + & \mu C\otimes\left(|0,0,1\rangle\langle 0,0,0|_u + |1,1,0\rangle\langle 1,1,1|_u\right)\nonumber
\end{eqnarray}
and $V_{-+} = \left(V_{+-}\right)^\dagger$.
A detailed calculation of these terms is provided in the Appendix. Substituting these values into Eq.~\ref{eff} we see that at order $k = 3$ a term appears with the following form,
\begin{equation}
\frac{1}{\Delta^2} V_{-+} V_+ V_{+-} = \frac{\mu^3}{\Delta^2} \left(A B C + A C B + B C A + C A B + B A C + CBA\right) \rightarrow A B C.
\end{equation}
These terms arise because all ancilla qubits must be flipped and there are six ways of doing so, representing $3!$ (in general this will be $k!$ for a gadget with $k$ ancillae) combinations of the operators. These six terms are represented diagrammatically in Figure~\ref{sixfig}. Note that it is the occurrence of all orderings of the operators $A$, $B$ and $C$ that imposes the requirement that these operators commute. Accordingly, in order to realize our desired term we see that $\mu = \sqrt[k]{\frac{\Delta^{k-1}}{k!}}$. A few competing processes occur which contribute unwanted terms but these terms either vanish with increasing spectral gap $\Delta$, or they can be removed exactly by introducing terms into the compensation term $\Lambda$. The most straightforward way to compute $\Lambda$ is to evaluate the perturbation series to order $k$ and choose $\Lambda$ so that problematic terms disappear.
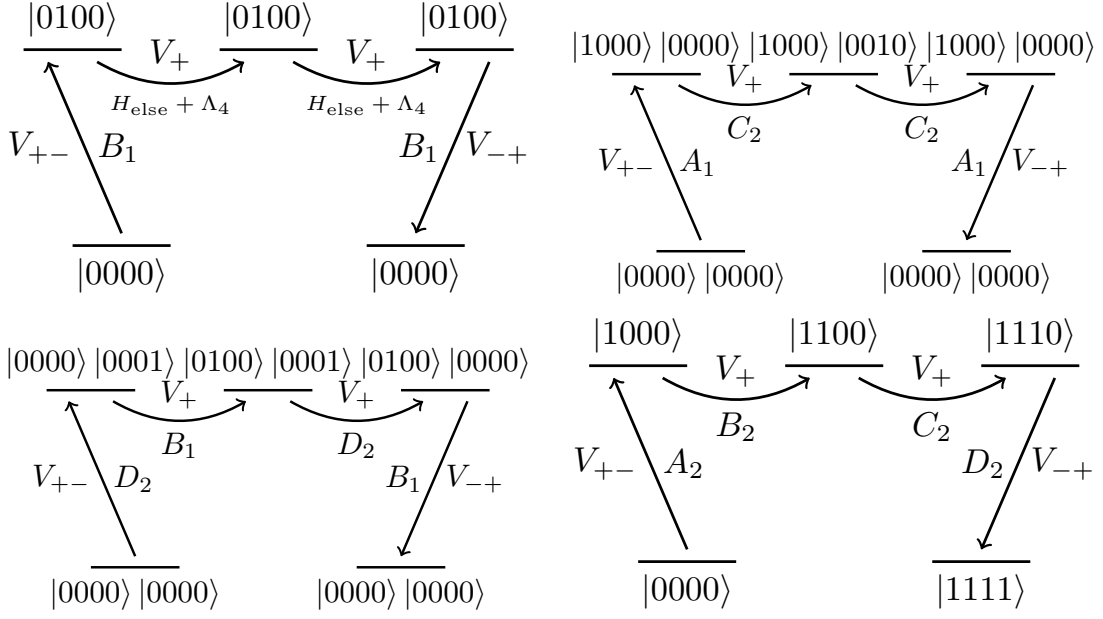
\begin{figure}[h]
\begin{center}
\begin{subfigure}[b]{.45\textwidth}
\resizebox{\linewidth}{!}{
    \begin{tikzpicture}[
      scale=0.5,
      level/.style={thick},
      trans/.style={->, thick},
      plus/.style={->, bend right, thick},
    ]
    \draw[level] (-4,0) -- (-2,0) node[midway, below] {$\ket{0000}$};
    \draw[level] (-5,4) -- (-3,4) node[midway, above] {$\ket{0100}$};
    \draw[level] (-1,4) -- (1,4) node[midway, above] {$\ket{0100}$};
    \draw[level] (3,4) -- (5,4) node[midway, above] {$\ket{0100}$};
    \draw[level] (2,0) -- (4,0) node[midway, below] {$\ket{0000}$};
    \draw[trans] (-3,0.25) to node[left] {$\small V_{+-}$} node[right] {\small $B_1$} (-4.5,3.75);
    \draw[plus] (-3.5,3.75) to node[above] {$\small V_{+}$} node[below] {\tiny $H_\textrm{else} + \Lambda_4$} (-.5,3.75);
    \draw[plus] (.5,3.75) to node[above] {$\small V_{+}$} node[below] {\tiny $H_\textrm{else} + \Lambda_4$} (3.5,3.75);
    \draw[trans] (4.5,3.75) to node[right] {\small $V_{-+}$}  node[left] {\small $B_1$} (3,0.25);
    \end{tikzpicture}}
\end{subfigure}
\begin{subfigure}[b]{.45\textwidth}
\resizebox{\linewidth}{!}{
    \begin{tikzpicture}[
      scale=0.5,
      level/.style={thick},
      trans/.style={->, thick},
      plus/.style={->, bend right, thick},
    ]
    \draw[level] (-4,0) -- (-2,0) node[midway, below] {\small $\ket{0000} \ket{0000}$};
    \draw[level] (-5,4) -- (-3,4) node[midway, above] {\small $\ket{1000}\ket{0000}$};
    \draw[level] (-1,4) -- (1,4) node[midway, above] {\small $\ket{1000}\ket{0010}$};
    \draw[level] (3,4) -- (5,4) node[midway, above] {\small$\ket{1000}\ket{0000}$};
    \draw[level] (2,0) -- (4,0) node[midway, below] {\small$\ket{0000}\ket{0000}$};
    \draw[trans] (-3,0.25) to node[left] {$\small V_{+-}$} node[right] {\small $A_1$} (-4.5,3.75);
    \draw[plus] (-3.5,3.75) to node[above] {$\small V_{+}$} node[below] {\small $C_2$} (-.5,3.75);
    \draw[plus] (.5,3.75) to node[above] {$\small V_{+}$} node[below] {\small $C_2$} (3.5,3.75);
    \draw[trans] (4.5,3.75) to node[right] {\small $V_{-+}$}  node[left] {\small $A_1$} (3,0.25);
    \end{tikzpicture}}
\end{subfigure}
\begin{subfigure}[b]{.45\textwidth}
\resizebox{\linewidth}{!}{
    \begin{tikzpicture}[
      scale=0.5,
      level/.style={thick},
      trans/.style={->, thick},
      plus/.style={->, bend right, thick},
    ]
    \draw[level] (-4,0) -- (-2,0) node[midway, below] {\small $\ket{0000} \ket{0000}$};
    \draw[level] (-5,4) -- (-3,4) node[midway, above] {\small $\ket{0000}\ket{0001}$};
    \draw[level] (-1,4) -- (1,4) node[midway, above] {\small $\ket{0100}\ket{0001}$};
    \draw[level] (3,4) -- (5,4) node[midway, above] {\small$\ket{0100}\ket{0000}$};
    \draw[level] (2,0) -- (4,0) node[midway, below] {\small$\ket{0000}\ket{0000}$};
    \draw[trans] (-3,0.25) to node[left] {$\small V_{+-}$} node[right] {\small $D_2$} (-4.5,3.75);
    \draw[plus] (-3.5,3.75) to node[above] {$\small V_{+}$} node[below] {\small $B_1$} (-.5,3.75);
    \draw[plus] (.5,3.75) to node[above] {$\small V_{+}$} node[below] {\small $D_2$} (3.5,3.75);
    \draw[trans] (4.5,3.75) to node[right] {\small $V_{-+}$}  node[left] {\small $B_1$} (3,0.25);
    \end{tikzpicture}}
\end{subfigure}
\begin{subfigure}[b]{.45\textwidth}
\resizebox{\linewidth}{!}{
    \begin{tikzpicture}[
      scale=0.5,
      level/.style={thick},
      trans/.style={->, thick},
      plus/.style={->, bend right, thick},
    ]
    \draw[level] (-4,0) -- (-2,0) node[midway, below] {$\ket{0000}$};
    \draw[level] (-5,4) -- (-3,4) node[midway, above] {$\ket{1000}$};
    \draw[level] (-1,4) -- (1,4) node[midway, above] {$\ket{1100}$};
    \draw[level] (3,4) -- (5,4) node[midway, above] {$\ket{1110}$};
    \draw[level] (2,0) -- (4,0) node[midway, below] {$\ket{1111}$};
    \draw[trans] (-3,0.25) to node[left] {$\small V_{+-}$} node[right] {\small $A_2$} (-4.5,3.75);
    \draw[plus] (-3.5,3.75) to node[above] {$\small V_{+}$} node[below] {\small $B_2$} (-.5,3.75);
    \draw[plus] (.5,3.75) to node[above] {$\small V_{+}$} node[below] {\small $C_2$} (3.5,3.75);
    \draw[trans] (4.5,3.75) to node[right] {\small $V_{-+}$}  node[left] {\small $D_2$} (3,0.25);
    \end{tikzpicture}}
\end{subfigure}
\end{center}
\caption{\label{crossgad}Diagrams showing an example of each of the four processes at fourth order. In the upper left is the process $B_1 \left(H_\textrm{else} + \Lambda\right)^2 B_1$. In the upper right is the process $A_1 C_2^2 A_1$. In the lower left is the process $D_2 B_1 D_2 B_1$. In the lower right is the process $A_2 B_2 C_2 D_2$.}
\end{figure}

At higher orders we encounter ``cross-gadget contamination'' which means that processes occur involving multiple ancilla systems, causing operators from different terms to interact. For a 3-operator gadget, such terms will always only contribute at order $O\left(\Delta^{-3}\right)$. In reductions which require going to higher orders, these terms do not necessarily depend on $\Delta$, and so may introduce unwanted terms into the effective Hamiltonian. For instance, Figure~\ref{crossgad} shows an example of the four processes which occur at fourth order for a multiple term, 4-operator reduction. The diagrams involving multiple ancilla registers are examples of cross-gadget contamination and do not disappear in the limit of large $\Delta$.

However, if terms are factored into tensor products of operators that square to the identity (as is the case for  Pauli operators, which we can always choose), cross-gadget contamination can only contribute a constant shift to the energy which may be easily compensated for in $\Lambda$. This must be true because any process contributing to the perturbation series which does not transition between the two different ground states must contain an even multiple of each operator and if we choose to act on the non-ancilla qubits with operators that square to identity we obtain only a constant shift. For instance, consider the two cross-gadget terms represented in these diagrams: $A_1 C_2 C_2 A_1 = A_1 \openone A_1 = \openone$ and $D_2  B_1 D_2 B_1 = \left(D_2 B_1\right)^2 = \openone$. At even higher orders, \emph{individual} cross-gadget terms might not equal a constant shift (i.e. the sixth order term $A_1 A_2 A_3 A_2 A_1 A_3$ but the occurrence of all combinations of operators and the fact that all Pauli terms either commute or anti-commute will guarantee that such terms disappear. For instance, in the sixth order example, if $\left[A_1, A_2\right] = 0$ then $A_1 A_2 A_3 A_2 A_1 A_3 = A_1 A_2 A_3 A_1 A_2 A_3 = \left(A_1 A_2 A_3\right)^2 = \openone$ otherwise we know that $\left[A_1, A_2\right]_+ = 0$ which implies that $A_1 A_2 A_3 A_2 A_1 A_3 + A_1 A_2 A_3 A_1 A_2 A_3 = 0$.

\section{Example Problem: Molecular Hydrogen}\label{sec:h2}

We begin by factoring and rewriting the $k$-local molecular hydrogen Hamiltonian from Eq.~\ref{H2} into a 2-local part and a $k$-local part so that $H_{\textrm{H}_2}  = H_\textrm{2-local} + H_\textrm{$4$-local}$ where,
\begin{eqnarray}
H_\textrm{2-local} & = & f_0 \openone +  f_2 Z_1 + f_3 Z_2 + f_4 Z_0 Z_2 +  f_5 Z_1 Z_3 + f_1 Z_0 \left(\openone + Z_1 \right)\\
H_\textrm{$4$-local} & = & \left(f_4 Z_0 + f_3 Z_1\right) Z_2 Z_3 + \left(Z_1 + Z_1 Z_3\right)\left(f_6 X_0 X_2 +  f_6 Y_0 Y_2 + f_7 Z_0 Z_2\right).
\end{eqnarray}
One could also divide the Hamiltonian into $2$, $3$, and $4$-local terms but this is less efficient than the procedure defined below. We focus on reducing $H_{\textrm{H}_2} $ to a 2-local $ZZ/XX/XZ$-Hamiltonian. We further factor $H_\textrm{4-local}$ to remove $YY$ terms,
\begin{eqnarray}
\label{firstred}
H_\textrm{$4$-local} & = & \underbrace{\left(f_4 Z_0 + f_3 Z_1\right)}_{A_1} \underbrace{Z_2}_{B_1} \underbrace{Z_3}_{C_1} + \underbrace{f_7 Z_0}_{A_2} \underbrace{Z_2}_{B_2} \underbrace{\left(Z_1 + Z_1 Z_3\right)}_{C_2} +  \underbrace{f_6 X_0 X_2}_{A_3} \underbrace{\left(\openone - Z_0 Z_2 \right)}_{B_3} \underbrace{\left(Z_1 + Z_1 Z_3\right)}_{C_3}\nonumber\\
& = & A_1 B_1 C_1 + A_2 B_2 C_2 + A_3 B_3 C_3.
\end{eqnarray}
Within each term, the operators all commute so that $\left[A_i,B_i\right]=\left[A_i,C_i\right]=\left[B_i,C_i\right]=0$. We emphasize that factoring terms into commuting operators is always possible and necessary in order for bit-flip gadgets to work correctly.

Each of the logical operators defined in Eq.~\ref{firstred} will have an associated ancilla qubit, e.g. the ancilla for operator $B_2$ has label $b_2$. Our unperturbed Hamiltonian is a sum of fully connected ancilla systems in which each ancilla system corresponds to a term,
\begin{eqnarray}
H_1 & = & \frac{9 \Delta_1}{4} \openone - \frac{\Delta_1}{4}\left(Z_{a_1}Z_{b_1} + Z_{a_1}Z_{c_1} + Z_{b_1}Z_{c_1} + Z_{a_2} Z_{b_2}\right.\\
& + & \left. Z_{a_2}Z_{c_2} + Z_{b_2}Z_{c_2} + Z_{a_3}Z_{b_3} + Z_{a_3}Z_{c_3} + Z_{b_3}Z_{c_3}\right)\nonumber.
\end{eqnarray}
The spectral gap and Hamiltonian have the subscript ``1'' to associate them with the first of two applications of perturbation theory. We perturb the ancilla system with the Hamiltonian,
\begin{eqnarray}
V_1 & = & \mu_1\left(A_1 X_{a_1} + B_1 X_{b_1} + C_1 X_{c_1} + A_2 X_{a_2}\right.\\
& + & \left. B_2 X_{b_2} + C_2 X_{c_2} + A_3 X_{a_3} + B_3 X_{b_3} + C_3 X_{c_3}\right) + H_\textrm{2-local} + \Lambda_1\nonumber
\end{eqnarray}
where $\mu_1 = \sqrt[3]{\frac{\Delta_1^2}{6}}$ and $\Lambda_1$ is a 2-local compensation Hamiltonian acting on the logical qubits only. Later on, $\Lambda_1$ will be chosen to cancel extraneous terms from the perturbative expansion. The interaction terms involving $A$, $B$, and $C$ will arise at third order ($V_{-+} V_+ V_{+-} $) from processes which involve a transition between the two degenerate ground states of the ancilla systems. This occurs at third order because to make the transition $|000\rangle \rightleftharpoons |111\rangle$, we must flip all three ancilla bits in each term by applying the operators $X_a$, $X_b$, and $X_c$. Since these operators are coupled to  $A$, $B$, and $C$, sequential action of bit flip operators yields our desired term. Because the operators commute, the order of the bit flips does not matter. We now calculate the effective Hamiltonian using the perturbative expansion of the self-energy from Eq.~\ref{eff}.

\subsection{Second Order}
The only processes which start in the ground state and return to the ground state at second order are those which flip a single bit and then flip the same bit back. Thus, effective interactions are created between each operator and itself,
\begin{eqnarray}
-\frac{1}{\Delta_1}V_{-+} V_{+-} & = & -\frac{\mu_1^2}{\Delta_1}\left(A_1^2 + B_1^2 + C_1^2 + A_2^2 + B_2^2 + C_2^2 + A_3^2 + B_3^2 + C_3^2\right)\\
& = & -\sqrt[3]{\frac{\Delta_1}{36}}\left[\left(9 + f_3^2 + f_4^2 + f_6^2 + f_7^2\right)\openone + 2 f_3 f_4 Z_0 Z_1 - 2 Z_0 Z_2 + 4 Z_3\right]\nonumber.
\end{eqnarray}
These processes are shown in Figure~\ref{2ndO}. The second order effective Hamiltonian at large $\Delta_1$ is,
\begin{equation}
H_\textrm{eff}^{\left(2\right)} = H_\textrm{2-local} + \Lambda_1 - \sqrt[3]{\frac{\Delta_1}{36}}\left[\left(9 + f_3^2 + f_4^2 + f_6^2 + f_7^2\right)\openone + 2 f_3 f_4 Z_0 Z_1 - 2 Z_0 Z_2 + 4 Z_3\right] + O\left(\Delta_1^{-2}\right).
\end{equation}
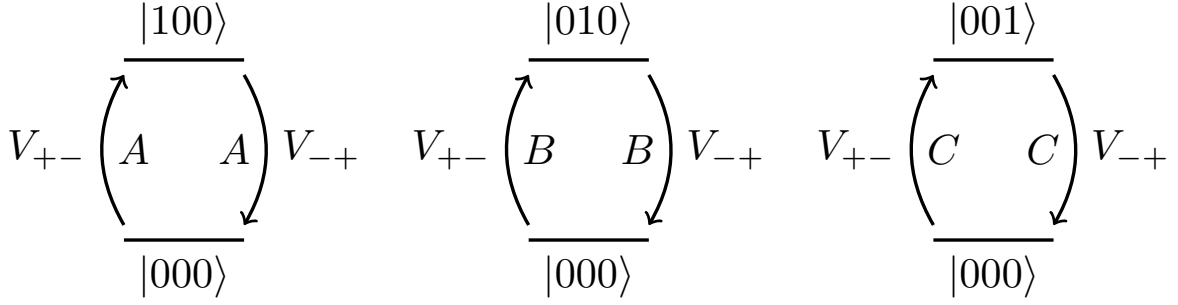
\begin{figure}[h]
\centering
\begin{subfigure}[b]{0.32\textwidth}
\resizebox{\linewidth}{!}{
    \begin{tikzpicture}[
      scale=0.5,
      level/.style={thick},
      trans/.style={->, thick},
      plus/.style={->, bend right, thick},
    ]
    \draw[level] (0,0) -- (2,0) node[midway, below] {$\ket{000}$};
    \draw[level] (0,3) -- (2,3) node[midway, above] {$\ket{100}$};
    \draw[->,bend left, thick] (0,0.25) to node[left] {$V_{+-}$} node[right] {$A$} (0,2.75);
    \draw[->,bend left, thick] (2,2.75) to node[left] {$A$} node[right] {$V_{-+}$} (2,.25);
    \end{tikzpicture}}
\end{subfigure}
\begin{subfigure}[b]{0.32\textwidth}
\resizebox{\linewidth}{!}{
    \begin{tikzpicture}[
      scale=0.5,
      level/.style={thick},
      trans/.style={->, thick},
      plus/.style={->, bend right, thick},
    ]
    \draw[level] (0,0) -- (2,0) node[midway, below] {$\ket{000}$};
    \draw[level] (0,3) -- (2,3) node[midway, above] {$\ket{010}$};
    \draw[->,bend left, thick] (0,0.25) to node[left] {$V_{+-}$} node[right] {$B$} (0,2.75);
    \draw[->,bend left, thick] (2,2.75) to node[left] {$B$} node[right] {$V_{-+}$} (2,.25);
    \end{tikzpicture}}
\end{subfigure}
\begin{subfigure}[b]{0.32\textwidth}
\resizebox{\linewidth}{!}{
    \begin{tikzpicture}[
      scale=0.5,
      level/.style={thick},
      trans/.style={->, thick},
      plus/.style={->, bend right, thick},
    ]
    \draw[level] (0,0) -- (2,0) node[midway, below] {$\ket{000}$};
    \draw[level] (0,3) -- (2,3) node[midway, above] {$\ket{001}$};
    \draw[->,bend left, thick] (0,0.25) to node[left] {$V_{+-}$} node[right] {$C$} (0,2.75);
    \draw[->,bend left, thick] (2,2.75) to node[left] {$C$} node[right] {$V_{-+}$} (2,.25);
    \end{tikzpicture}}
\end{subfigure}
\caption{\label{2ndO}The three bit-flip processes at second order. These occur for each term. Note that each of these diagrams occurs in reverse for the part of the ground state in $\ket{111}$.}
\end{figure}

\subsection{Third Order}
\begin{figure}[h]
\centering
\begin{subfigure}[n]{.32\textwidth}
\resizebox{\linewidth}{!}{
    \begin{tikzpicture}[
      scale=0.5,
      level/.style={thick},
      trans/.style={->, thick},
      plus/.style={->, bend right, thick},
    ]
    \draw[level] (1,0) -- (3,0) node[midway, below] {$\ket{000}$};
    \draw[level] (0,4) -- (2,4) node[midway, above] {$\ket{100}$};
    \draw[level] (5,4) -- (7,4) node[midway, above] {$\ket{100}$};
    \draw[level] (4,0) -- (6,0) node[midway, below] {$\ket{000}$};
    \draw[trans] (2,0.25) to node[left] {$V_{+-}$} node[right] {$A$} (0.5,3.75);
    \draw[plus] (1.5,3.75) to node[above] {$V_{+}$} node[below] {\tiny $H_\textrm{2-local} + \Lambda_1$} (5.5,3.75);
    \draw[trans] (6.5,3.75) to node[right] {$V_{-+}$}  node[left] {$A$} (5,0.25);
    \end{tikzpicture}}
\end{subfigure}
\begin{subfigure}[n]{.32\textwidth}
\resizebox{\linewidth}{!}{
    \begin{tikzpicture}[
      scale=0.5,
      level/.style={thick},
      trans/.style={->, thick},
      plus/.style={->, bend right, thick},
    ]
    \draw[level] (1,0) -- (3,0) node[midway, below] {$\ket{000}$};
    \draw[level] (0,4) -- (2,4) node[midway, above] {$\ket{010}$};
    \draw[level] (5,4) -- (7,4) node[midway, above] {$\ket{010}$};
    \draw[level] (4,0) -- (6,0) node[midway, below] {$\ket{000}$};
    \draw[trans] (2,0.25) to node[left] {$V_{+-}$} node[right] {$B$} (0.5,3.75);
    \draw[plus] (1.5,3.75) to node[above] {$V_{+}$} node[below] {\tiny $H_\textrm{2-local} + \Lambda_1$} (5.5,3.75);
    \draw[trans] (6.5,3.75) to node[right] {$V_{-+}$}  node[left] {$B$} (5,0.25);
    \end{tikzpicture}}
\end{subfigure}
\begin{subfigure}[n]{.32\textwidth}
\resizebox{\linewidth}{!}{
    \begin{tikzpicture}[
      scale=0.5,
      level/.style={thick},
      trans/.style={->, thick},
      plus/.style={->, bend right, thick},
    ]
    \draw[level] (1,0) -- (3,0) node[midway, below] {$\ket{000}$};
    \draw[level] (0,4) -- (2,4) node[midway, above] {$\ket{001}$};
    \draw[level] (5,4) -- (7,4) node[midway, above] {$\ket{001}$};
    \draw[level] (4,0) -- (6,0) node[midway, below] {$\ket{000}$};
    \draw[trans] (2,0.25) to node[left] {$V_{+-}$} node[right] {$C$} (0.5,3.75);
    \draw[plus] (1.5,3.75) to node[above] {$V_{+}$} node[below] {\tiny $H_\textrm{2-local} + \Lambda_1$} (5.5,3.75);
    \draw[trans] (6.5,3.75) to node[right] {$V_{-+}$}  node[left] {$C$} (5,0.25);
    \end{tikzpicture}}
\end{subfigure}
\caption{\label{3rdO}Diagrams for the competing process encountered at third order. Note that each of these diagrams can also occur backwards if the system starts in $\ket{111}$.}
\end{figure}
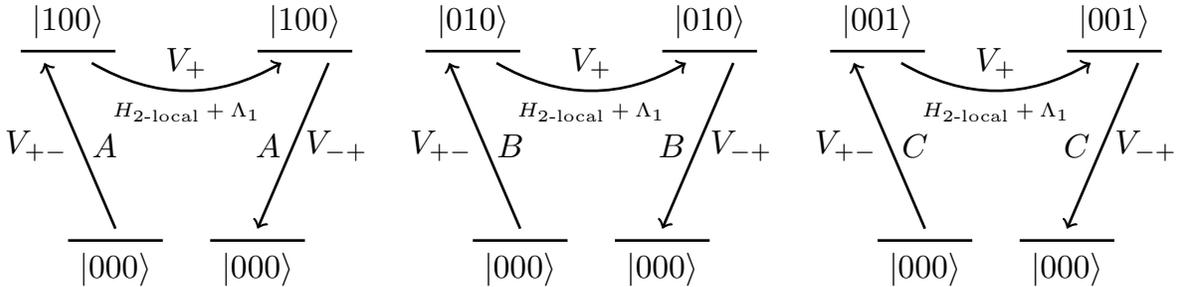

The target Hamiltonian terms appears at third order from processes that transition between degenerate ground states. However, there is also an additional, unwanted process which occurs at this order. This competing process involves one interaction with $H_\textrm{2-local}$ and $\Lambda_1$ in the high-energy subspace,
\begin{eqnarray}
\label{3,1}
\frac{1}{\Delta_1^2} V_{-+} & V_+ & V_{+-}^{\left(1\right)} = \frac{\mu_1^2}{\Delta_1^2}\left[A_1\left(H_\textrm{2-local} + \Lambda_1\right)  A_1 + B_1 \left(H_\textrm{2-local} + \Lambda_1\right) B_1 + C_1\left(H_\textrm{2-local} + \Lambda_1\right) C_1 \right.\\
& + & \left. A_2\left(H_\textrm{2-local} + \Lambda_1\right)  A_2 + B_2\left(H_\textrm{2-local} + \Lambda_1\right)  B_2 + C_2 \left(H_\textrm{2-local} + \Lambda_1\right) C_2 \right.\nonumber\\
& + & \left. A_3 \left(H_\textrm{2-local} + \Lambda_1\right) A_3 +  B_3 \left(H_\textrm{2-local} + \Lambda_1\right) B_3 + C_3 \left(H_\textrm{2-local} + \Lambda_1\right) C_3\right]\nonumber.
\end{eqnarray}
These processes are illustrated diagrammatically in Figure~\ref{3rdO}.
The process we want occurs with the ancilla transition $|000\rangle \rightleftharpoons |111\rangle$ which flips all three bits (for each term separately since they have different ancillae). There are $3! = 6$ possible ways to flip the bits for each term, as illustrated in Figure~\ref{sixfig} of Section~\ref{sec:g1}. This gives us the term,
\begin{equation}
\label{3,2}
\frac{1}{\Delta_1^2}V_{-+} V_+ V_{+-}^{\left(2\right)} = 6 \frac{\mu_1^3}{\Delta_1^2} \left( A_1 B_1 C_1 + A_2 B_2 C_2  + A_3 B_3 C_3 \right) = A_1 B_1 C_1 + A_2 B_2 C_2 + A_3 B_3 C_3.
\end{equation}

Because $H_\textrm{2-local}$ has no $\Delta_1$ dependence and $\mu_1$ is order $O\left(\Delta_1^{2/3}\right)$, terms such as $\left( \mu_1^2 / \Delta_1^2\right) A_1 H_\textrm{2-local} A_1$ will vanish in the limit of large $\Delta_1$; therefore, the third order effective Hamiltonian is,
\begin{eqnarray}
\label{3,1}
H_\textrm{eff}^{\left(3\right)} & = & H_\textrm{2-local} + \Lambda_1 -\sqrt[3]{\frac{\Delta_1}{36}}\left[\left(9 + f_3^2 + f_4^2 + f_6^2 + f_7^2\right)\openone + 2 f_3 f_4 Z_0 Z_1 - 2 Z_0 Z_2 + 4 Z_3\right]\\
& + & \frac{\mu_1^2}{\Delta_1^2}\left(A_1 \Lambda_1 A_1 + B_1 \Lambda_1 B_1 + C_1 \Lambda_1 C_1 + A_2 \Lambda_1 A_2 + B_2 \Lambda_1 B_2 + C_2 \Lambda_1 C_2\right.\nonumber\\
& + & \left. A_3 \Lambda_1 A_3 +  B_3  \Lambda_1 B_3 + C_3 \Lambda_1 C_3 \right) + A_1 B_1 C_1 + A_2 B_2 C_2 + A_3 B_3 C_3 + O\left(\Delta_1^{-3}\right).\nonumber
\end{eqnarray}
By inspection we see that if $\Lambda_1 = \frac{1}{\Delta_1}V_{-+} V_{+-}$ then the unwanted contribution at third order will go to zero in the limit of large $\Delta_1$ and the second order term will cancel exactly with $\Lambda_1$ leaving,
\begin{eqnarray}
H_\textrm{eff}^{\left(3\right)} & = & H_\textrm{2-local} + A_1 B_1 C_1 + A_2 B_2 C_2 + A_3 B_3 C_3 + O\left(\Delta_1^{-3}\right)\\
H_{\textrm{H}_2} & \rightarrow & H_1 + V_1,
\end{eqnarray}
where ``$\rightarrow$'' denotes an embedding. There are still 3-local terms remaining in $V_1$,
\begin{eqnarray}
V_1 & = & \mu_1 \left(f_4 Z_0 + f_3 Z_1\right) X_{a_1} + \mu_1 X_2 \left(X_{b_1} +X_{b_2}\right) + \mu_1 Z_3 X_{c_1} + \mu_1f_7 Z_0 X_{a_2} + \mu_1 Z_1 \left(Z_{c_2} + X_{c_3}\right)\,\,\,\,\,\\
& + & \underbrace{\mu_1 Z_1}_{A_4} \underbrace{Z_3}_{B_4} \underbrace{\left(X_{c_2} + X_{c_3}\right)}_{C_4} + \underbrace{\mu_1 f_6 X_0}_{A_5} \underbrace{X_2}_{B_5} \underbrace{X_{a_3}}_{C_5} + \mu_1 X_{b_3} + \underbrace{\left(-\mu_1\right)Z_0}_{A_6}\underbrace{Z_2}_{B_6}\underbrace{X_{b_3}}_{C_6} +  H_\textrm{2-local}  + \Lambda_1 \nonumber.
\end{eqnarray} 

\begin{figure}[H]
\centering
\includegraphics[scale = .45]{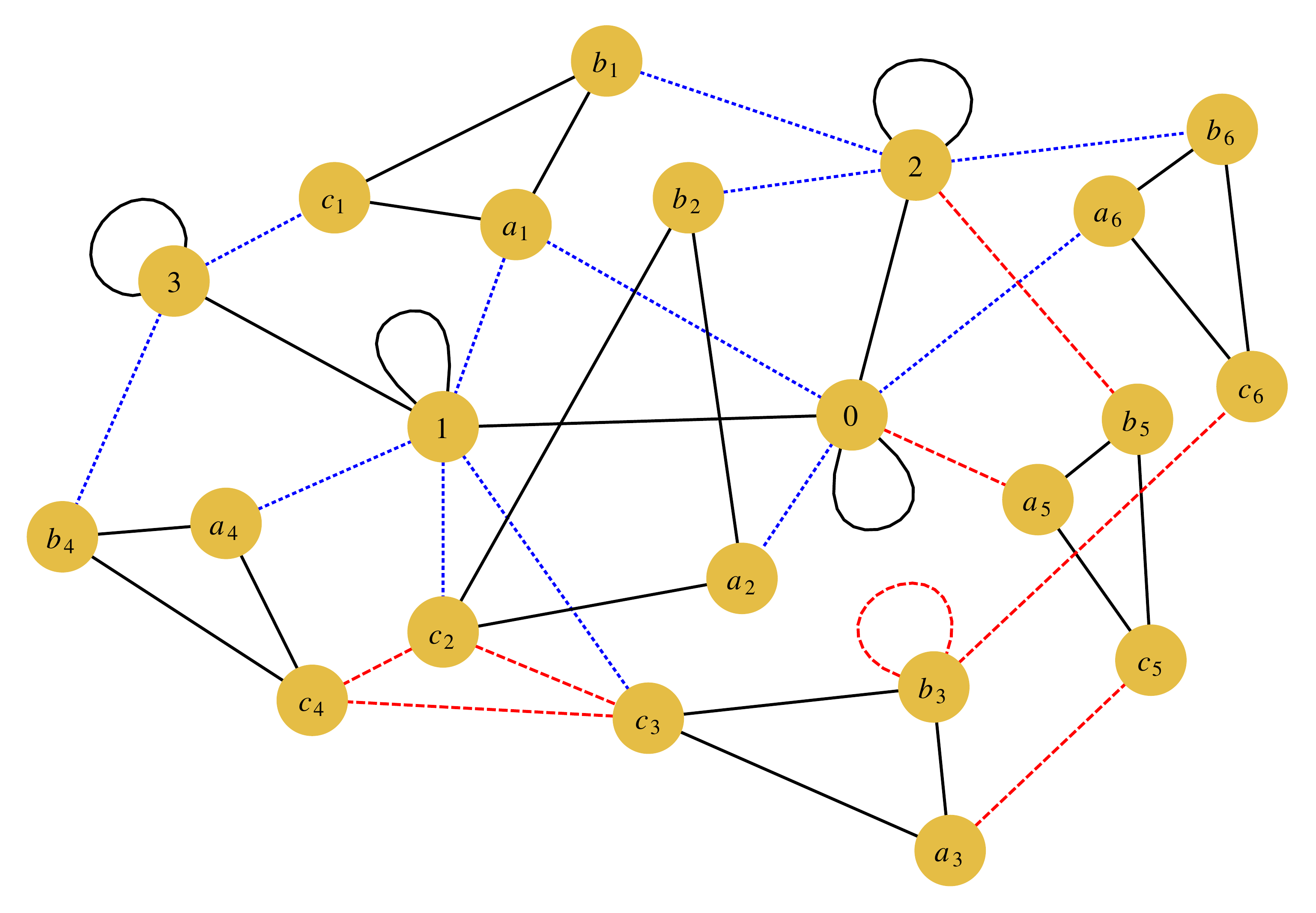}
\caption{\label{interaction}Interaction graph for embedded molecular hydrogen Hamiltonian. Each node represents a qubit. The solid, black edges represent local $Z$ or $ZZ$ terms. The dashed, red edges represent local $X$ or $XX$ terms. The dotted, blue edges represent $XZ$ terms. It is easy to see the unperturbed Hamiltonians corresponding to the six 3-operator terms (the black triangles).}
\end{figure}

With this notation we reorganize our Hamiltonian a final time,
so that $H_{\textrm{H}_2} \rightarrow H_\textrm{2-local} + H_\textrm{3-local}$,
\begin{eqnarray}
H_\textrm{2-local} & = &  \left(f_0 + \frac{9 \Delta_1}{4}\right) \openone +  f_2 Z_1+ f_3 Z_2 + f_4 Z_0 Z_2 +  f_5 Z_1 Z_3 + f_1 Z_0  \left(\openone + Z_1\right)\\
& - & \frac{\Delta_1}{4}\left(Z_{a_1} Z_{b_1} + Z_{a_1} Z_{c_1} + Z_{b_1} Z_{c_1} + Z_{a_2} Z_{b_2} + Z_{a_2} Z_{c_2} + Z_{b_2} Z_{c_2} + Z_{a_3}Z_{b_3} + Z_{a_3} Z_{c_3} + Z_{b_3} Z_{c_3}\right)\nonumber\\
& + & \sqrt[3]{\frac{\Delta_1^2}{6}} \left[\left(f_4 Z_0 + f_3 Z_1\right) X_{a_1} + Z_3 X_{c_1} + X_2 \left(X_{b_1} + X_{b_2}\right) +f_7 Z_0 X_{a_2} + Z_1 \left(X_{c_2} + X_{c_3}\right) + X_{b_3} \right]\nonumber\\
& + & \sqrt[3]{\frac{\Delta_1}{36}}\left[\left(9 + f_3^2 + f_4^2 + f_6^2 + f_7^2\right)\openone + 2 f_3 f_4 Z_0 Z_1 - 2 Z_0 Z_2 + 4 Z_3\right]\nonumber\\
H_\textrm{3-local} & = & A_4 B_4 C_4 + A_5 B_5 C_5 + A_6 B_6 C_6.
\end{eqnarray}
The third order gadget we need to reduce $H_\textrm{3-local}$ takes exactly the same form as before except with the term labels $1,2,3$ exchanged for the term labels $4,5,6$. The gadget Hamiltonian is
\begin{eqnarray}
H_2 & = & \frac{9 \Delta_2}{4} \openone - \frac{\Delta_2}{4}\left(Z_{a_4} Z_{b_4} + Z_{a_4} Z_{c_4} + Z_{b_4} Z_{c_4} + Z_{a_5} Z_{b_5}\right.\\
& + & \left. Z_{a_5} Z_{c_5} + Z_{b_5} Z_{c_5} + Z_{a_6} Z_{b_6} + Z_{a_6} Z_{c_6} + Z_{b_6} Z_{c_6}\right)\nonumber\\
V_2 & = & \mu_2\left(A_4 X_{a_4} + B_4 X_{b_4} + C_4 X_{c_4} + A_5 X_{a_5}\right.\\
& + & \left. B_5 X_{b_5} + C_5 X_{c_5} + A_6 X_{a_6} + B_6 X_{b_6} + C_6 X_{c_6}\right) + H_\textrm{2-local} + \Lambda_2\nonumber
\end{eqnarray}
where $\mu_2 = \sqrt[3]{\frac{\Delta_2^2}{6}}$ and
\begin{eqnarray}
 \Lambda_2 & = & \frac{\mu_2^2}{\Delta_2}\left(A_4^2 + B_4^2 + C_4^2 + A_5^2 + B_5^2 + C_5^2 + A_6^2 + B_6^2 + C_6^2\right)\\
& = & \left(7 \sqrt[3]{\frac{\Delta_2}{36}} + \frac{\Delta_1^{4/3}}{3} \sqrt[3]{\frac{\Delta_2}{6}} +  \frac{f_6^2 \Delta_1^{4/3}}{6} \sqrt[3]{\frac{\Delta_2}{6}}\right)\openone +  \sqrt[3]{\frac{2 \Delta_2}{9}} X_{c_2} X_{c_3} \nonumber.
\end{eqnarray}
This time the spectral gap and Hamiltonian have the subscript ``2'' to associate them with our second application of perturbation theory. We have thus shown the embedding $H_{\textrm{H}_2} \rightarrow H_2 + V_2$. We present an interaction graph for the embedded Hamiltonian in Figure~\ref{interaction}.

\section{Conclusion}\label{sec:conc}

We have presented a fully general method for mapping any molecular electronic structure instance to a 2-local Hamiltonian containing only $ZZ$, $XX$ and $XZ$ terms. Our method is scalable in the sense that all experimental resources (qubits, control precision, graph degree) scale polynomially in the number of orbitals. We used perturbative gadgets which embed the entire target Hamiltonian (as opposed to just the ground state), thus guaranteeing that the eigenvalue gap is conserved under our reduction. Furthermore, we showed that bit-flip gadgets can be applied to remove experimentally challenging $YY$ terms. The resulting Hamiltonian is suitable for implementation in superconducting systems, quantum dots and other systems of artificial spins with the correct engineered interactions.

Further reduction of the types of interactions present is possible, to either $ZZ$ and $XX$ terms or $ZZ$ and $XZ$ terms, using the techniques of~\cite{Biamonte2007}. This makes the required interactions for simulating electronic structure Hamiltonians equivalent to the requirements of universal adiabatic quantum computation~\cite{Biamonte2007}. However, repeated reduction of the Hamiltonian results in more stringent precision requirements. The chosen target set of interactions strikes a balance between control precision and a reasonable set of distinct types of controllable interaction. The techniques developed here could also be applied to interacting fermion problems on the lattice. However, in that case it is possible to improve beyond the Bravyi-Kitaev mapping and exploit the locality of the interactions to directly obtain Hamiltonians whose locality is independent of the number of orbitals~\cite{Verstraete2005}.

We intend to follow-up this work with an analysis of hardware requirements for implementation on a system with superconducting qubits. A detailed scaling study of the exact resources needed for this algorithm as a function of molecular size is underway. We propose to read out energy eigenvalues using the tunneling spectroscopy of a probe qubit. This technique has already been demonstrated experimentally with rf SQUID flux qubits in \cite{Berkley2013}. In this scheme, a probe qubit is coupled to a single qubit of the simulation. Tunneling transitions allow the probe qubit to flip when the energy bias of the probe is close to an eigenvalue of the original system. Hence detection of  these transitions reveals the eigenspectrum of the original system. In this way, we would be able to directly measure the eigenspectra of the molecular systems embedded into the spin Hamiltonian using the techniques developed in the present paper.

There has been rapid recent progress in new classical algorithms, such as DMRG (density matrix renormalization group) and related tensor network methods, and proving complexity and approximability results pertaining to minimal resource model Hamiltonians. By using and understanding the techniques we have introduced in this paper, problems in chemistry can be  reduced to such models and these discoveries can be leveraged to make advances in electronic structure theory. However, we note that the spin Hamiltonians that result from the mapping developed here will be {\em non-stoquastic}, and classical simulation techniques will therefore suffer from the fermionic sign problem~\cite{Bravyi2006}. This further motivates the construction of quantum hardware to address the electronic structure problem by quantum simulation of these spin Hamiltonians.

\section{Acknowledgements}
The authors thank Dr. Sarah Mostame and Yudong Cao for helpful comments during revision. A. A.-G. acknowledges funding from National Science Foundation grant CHE-1152291 and the Air Force Office of Scientific Research under contract FA9550-12-1-0046. P.J.L. acknowledges National Science Foundation grant PHY-0955518. A.A.-G. and R. B. thank the Corning Foundation for their generous support. Research sponsored by the United States Department of Defense. The views and conclusions contained in this document are those of the authors and should not be interpreted as representing the official policies, either expressly or implied, of the United States Government.

\bibliographystyle{ieeetr}
\bibliography{library}

\begin{thebibliography}{100}

\bibitem{Farhi2000}
E.~Farhi, J.~Goldstone, S.~Gutmann, and M.~Sipser, ``{Quantum Computation by
  Adiabatic Evolution},'' {\em e-print arXiv:0001106}, 2000.

\bibitem{Farhi2001}
E.~Farhi, J.~Goldstone, S.~Gutmann, J.~Lapan, A.~Lundgren, and D.~Preda, ``{A
  Quantum Adiabatic Evolution Algorithm Applied to Random Instances of an
  NP-Complete Problem},'' {\em Science}, vol.~292, no.~5516, pp.~472--475,
  2001.

\bibitem{Born}
M.~Born and V.~Fock, ``{Beweis des Adiabatensatzes},'' {\em Zeitschrift f\"{u}r
  Physik A}, vol.~51, pp.~165--180, 1928.

\bibitem{Boixo2009}
S.~Boixo and R.~D. Somma, ``{Necessary Condition for the Quantum Adiabatic
  Approximation},'' {\em Physical Review A}, vol.~81, no.~3, p.~5, 2009.

\bibitem{GarveyJohnson}
M.~R. Garey and D.~S. Johnson, {\em {Computers and Intractability: A Guide to
  the Theory of NP-Completeness}}.
\newblock W. H. Freeman, 1979.

\bibitem{Hogg2003}
T.~Hogg, ``{Adiabatic quantum computing for random satisfiability problems},''
  {\em Physical Review A}, vol.~67, p.~22314, Feb. 2003.

\bibitem{Choi2010}
V.~Choi, ``{Adiabatic Quantum Algorithms for the NP-Complete Maximum-Weight
  Independent Set, Exact Cover and 3SAT Problems},'' {\em e-print
  arXiv:1004.2226}, p.~2226, Apr. 2010.

\bibitem{Nehaus2011}
T.~Neuhaus, M.~Peschina, K.~Michielsen, and H.~{De Raedt}, ``{Classical and
  quantum annealing in the median of three-satisfiability},'' {\em Physical
  Review A}, vol.~83, p.~12309, Jan. 2011.

\bibitem{Farhi2012}
E.~Farhi, D.~Gosset, I.~Hen, A.~Sandvik, A.~Young, P.~Shor, and F.~Zamponi,
  ``{Performance of the quantum adiabatic algorithm on random instances of two
  optimization problems on regular hypergraphs},'' {\em Physical Review A},
  vol.~86, no.~5, 2012.

\bibitem{Boixo2013}
S.~Boixo, T.~F. Ronnow, S.~V. Isakov, Z.~Wang, D.~Wecker, D.~A. Lidar, J.~M.
  Martinis, M.~Troyer, and I.~V. Sergei, ``{Quantum annealing with more than
  one hundred qubits},'' {\em e-print arxiv:1304.4595}, 2013.

\bibitem{Perdomo-Ortiz2012}
A.~Perdomo-Ortiz, N.~Dickson, M.~Drew-Brook, G.~Rose, and A.~Aspuru-Guzik,
  ``{Finding low-energy conformations of lattice protein models by quantum
  annealing},'' {\em Scientific Reports}, vol.~2, 2012.

\bibitem{Babbush2012}
R.~Babbush, A.~Perdomo-ortiz, B.~O. Gorman, W.~Macready, A.~Aspuru-Guzik, and
  B.~O'Gorman, ``{Construction of Energy Functions for Lattice Heteropolymer
  Models : A Case Study in Constraint Satisfaction Programming and Adiabatic
  Quantum Optimization},'' {\em e-print arXiv:1211.3422}, p.~42, Nov. 2012.

\bibitem{Denchev2012}
V.~S. Denchev, N.~Ding, S.~V.~N. Vishwanathan, and H.~Neven, ``{Robust
  Classification with Adiabatic Quantum Optimization},'' {\em Proc Int Conf on
  Machine Learning}, p.~1205.1148, 2012.

\bibitem{Neven2009}
H.~Neven, V.~S. Denchev, G.~Rose, and W.~G. Macready, ``{Training a Large Scale
  Classifier with the Quantum Adiabatic Algorithm},'' {\em e-print
  arXiv:09120779}, p.~14, 2009.

\bibitem{Roland:2002}
J.~Roland and N.~J. Cerf, ``{Quantum search by local adiabatic evolution},''
  {\em Physical Review A}, vol.~65, p.~42308, Mar. 2002.

\bibitem{Roland:2003}
J.~Roland and N.~J. Cerf, ``{Adiabatic quantum search algorithm for structured
  problems},'' {\em Physical Review A}, vol.~68, p.~62312, Dec. 2003.

\bibitem{Garnerone2012}
S.~Garnerone, P.~Zanardi, and D.~A. Lidar, ``{Adiabatic Quantum Algorithm for
  Search Engine Ranking},'' {\em Physical Review Letters}, vol.~108, p.~230506,
  June 2012.

\bibitem{Smelyanskiy2012}
V.~N. Smelyanskiy, E.~G. Rieffel, S.~I. Knysh, C.~P. Williams, M.~W. Johnson,
  M.~C. Thom, W.~G. Macready, and K.~L. Pudenz, ``{A Near-Term Quantum
  Computing Approach for Hard Computational Problems in Space Exploration},''
  {\em Electrical Engineering}, p.~68, 2012.

\bibitem{Barahona1982}
F.~Barahona, ``{On the computational complexity of Ising spin glass models},''
  {\em Journal of Physics A: Mathematical and General}, vol.~15, no.~10,
  p.~3241, 1982.

\bibitem{Bernstein:1997}
E.~Bernstein and U.~Vazirani, ``{Quantum complexity theory},'' {\em SIAM
  Journal on Computing}, vol.~26, no.~5, pp.~1411--1473, 1997.

\bibitem{Harris2007}
R.~Harris, A.~J. Berkley, M.~W. Johnson, P.~Bunyk, S.~Govorkov, M.~C. Thom,
  S.~Uchaikin, {Wilson, AB}, J.~Chung, and E.~Holtham, ``{Sign-and
  magnitude-tunable coupler for superconducting flux qubits},'' {\em Physical
  Review Letters}, vol.~98, no.~17, p.~177001, 2007.

\bibitem{Harris2008}
R.~Harris, M.~W. Johnson, S.~Han, A.~J. Berkley, J.~Johansson, P.~Bunyk,
  E.~Ladizinsky, S.~Govorkov, M.~C. Thom, S.~Uchaikin, B.~Bumble, A.~Fung,
  A.~Kaul, A.~Kleinsasser, M.~H.~S. Amin, and D.~V. Averin, ``{Probing Noise in
  Flux Qubits via Macroscopic Resonant Tunneling},'' {\em Physical Review
  Letters}, vol.~101, p.~117003, Sept. 2008.

\bibitem{Harris2009b}
R.~Harris, F.~Brito, A.~J. Berkley, J.~Johansson, M.~W. Johnson, T.~Lanting,
  P.~Bunyk, E.~Ladizinsky, B.~Bumble, and A.~Fung, ``{Synchronization of
  multiple coupled rf-SQUID flux qubits},'' {\em New Journal Of Physics},
  vol.~11, no.~12, p.~123022, 2009.

\bibitem{Harris2009c}
R.~Harris, T.~Lanting, A.~J. Berkley, J.~Johansson, M.~W. Johnson, P.~Bunyk,
  E.~Ladizinsky, N.~Ladizinsky, T.~Oh, and S.~Han, ``{Compound
  Josephson-junction coupler for flux qubits with minimal crosstalk},'' {\em
  Physical Review B}, vol.~80, no.~5, p.~52506, 2009.

\bibitem{Lanting2009}
T.~Lanting, A.~J. Berkley, B.~Bumble, P.~Bunyk, A.~Fung, J.~Johansson, A.~Kaul,
  A.~Kleinsasser, E.~Ladizinsky, and F.~Maibaum, ``{Geometrical dependence of
  the low-frequency noise in superconducting flux qubits},'' {\em Physical
  Review B}, vol.~79, no.~6, p.~60509, 2009.

\bibitem{Johansson2009}
J.~Johansson, M.~H.~S. Amin, A.~J. Berkley, P.~Bunyk, V.~Choi, R.~Harris, M.~W.
  Johnson, T.~M. Lanting, S.~Lloyd, and G.~Rose, ``{Landau-Zener transitions in
  a superconducting flux qubit},'' {\em Physical Review B}, vol.~80, p.~12507,
  July 2009.

\bibitem{Berkley2010}
A.~J. Berkley, M.~W. Johnson, P.~Bunyk, R.~Harris, J.~Johansson, T.~Lanting,
  E.~Ladizinsky, E.~Tolkacheva, M.~H.~S. Amin, and G.~Rose, ``{A scalable
  readout system for a superconducting adiabatic quantum optimization
  system},'' {\em Superconductor Science and Technology}, vol.~23, no.~10,
  p.~105014, 2010.

\bibitem{Berkley2013}
A.~J. Berkley, A.~J. Przybysz, T.~Lanting, R.~Harris, N.~Dickson, F.~Altomare,
  M.~H. Amin, P.~Bunyk, C.~Enderud, E.~Hoskinson, M.~W. Johnson, E.~Ladizinsky,
  R.~Neufeld, C.~Rich, A.~Y. Smirnov, E.~Tolkacheva, S.~Uchaikin, and A.~B.
  Wilson, ``{Tunneling spectroscopy using a probe qubit},'' {\em Physical
  Review B}, vol.~87, no.~2, p.~020502, 2013.

\bibitem{Johnson2011}
M.~W. Johnson, M.~H.~S. Amin, S.~Gildert, T.~Lanting, F.~Hamze, N.~Dickson,
  R.~Harris, A.~J. Berkley, J.~Johansson, P.~Bunyk, E.~M. Chapple, C.~Enderud,
  J.~P. Hilton, K.~Karimi, E.~Ladizinsky, N.~Ladizinsky, T.~Oh, I.~Perminov,
  C.~Rich, M.~C. Thom, E.~Tolkacheva, C.~J.~S. Truncik, S.~Uchaikin, J.~Wang,
  B.~Wilson, and G.~Rose, ``{Quantum annealing with manufactured spins.},''
  {\em Nature}, vol.~473, no.~7346, pp.~194--198, 2011.

\bibitem{Dickson2013}
N.~G. Dickson, M.~W. Johnson, M.~H. Amin, R.~Harris, F.~Altomare, a.~J.
  Berkley, P.~Bunyk, J.~Cai, E.~M. Chapple, P.~Chavez, F.~Cioata, T.~Cirip,
  P.~Debuen, M.~Drew-Brook, C.~Enderud, S.~Gildert, F.~Hamze, J.~P. Hilton,
  E.~Hoskinson, K.~Karimi, E.~Ladizinsky, N.~Ladizinsky, T.~Lanting, T.~Mahon,
  R.~Neufeld, T.~Oh, I.~Perminov, C.~Petroff, A.~Przybysz, C.~Rich, P.~Spear,
  A.~Tcaciuc, M.~C. Thom, E.~Tolkacheva, S.~Uchaikin, J.~Wang, a.~B. Wilson,
  Z.~Merali, and G.~Rose, ``{Thermally assisted quantum annealing of a 16-qubit
  problem.},'' {\em Nature communications}, vol.~4, no.~May, p.~1903, 2013.

\bibitem{Pudenz2013}
K.~L. Pudenz, T.~Albash, and D.~A. Lidar, ``{Error corrected quantum annealing
  with hundreds of qubits},'' {\em e-print arXiv:1307.8190}, p.~8190, July
  2013.

\bibitem{Bian2013}
Z.~Bian, F.~Chudak, W.~G. Macready, L.~Clark, and F.~Gaitan, ``{Experimental
  Determination of Ramsey Numbers},'' {\em Physical Review Letters}, vol.~111,
  p.~130505, Sept. 2013.

\bibitem{Wang2013}
L.~Wang, T.~F. Ronnow, S.~Boixo, S.~V. Isakov, Z.~Wang, D.~Wecker, D.~A. Lidar,
  J.~M. Martinis, and M.~Troyer, ``{Comment on: "Classical signature of quantum
  annealing".},'' {\em e-print arXiv:1305.5837}, 2013.

\bibitem{Smolin2013}
J.~A. Smolin and G.~Smith, ``{Classical signatures of quantum annealing},''
  {\em e-print arXiv:1305.4904}, 2013.

\bibitem{Shor:1997}
P.~W. Shor, ``{Polynomial-time algorithms for prime factorization and discrete
  logarithms on a quantum computer},'' {\em SIAM Journal on Computing},
  vol.~26, no.~5, pp.~1484--1509, 1997.

\bibitem{Childs:2003}
A.~M. Childs, R.~Cleve, E.~Deotto, E.~Farhi, S.~Gutmann, and D.~A. Spielman,
  ``{Exponential algorithmic speedup by a quantum walk},'' {\em Proceedings of
  the thirty-fifth annual ACM symposium on Theory of computing}, vol.~35,
  pp.~59--68, 2003.

\bibitem{Grover:1996}
L.~K. Grover, ``{A fast quantum mechanical algorithm for database search},'' in
  {\em Proceedings of the twenty-eighth annual ACM symposium on Theory of
  computing}, STOC '96, (New York, NY, USA), pp.~212--219, ACM, 1996.

\bibitem{Feynman1982}
R.~P. Feynman, ``{Simulating physics with computers},'' {\em International
  Journal of Theoretical Physics}, vol.~21, no.~6-7, pp.~467--488, 1982.

\bibitem{Meyer1996}
D.~A. Meyer, ``{From quantum cellular automata to quantum lattice gases},''
  {\em Journal of Statistical Physics}, vol.~85, no.~5-6, pp.~551--574, 1996.

\bibitem{Wiesner1996}
S.~Wiesner, ``{Simulations of many-body quantum systems by a quantum
  computer},'' {\em e-print arXiv:9603028}, vol.~110, 1996.

\bibitem{Lloyd1996}
S.~Lloyd, ``{Universal Quantum Simulators},'' {\em Science}, vol.~273,
  pp.~1073--1078, Aug. 1996.

\bibitem{Lidar1997}
D.~A. Lidar and O.~Biham, ``{Simulating Ising spin glasses on a quantum
  computer},'' {\em Physical Review E (Statistical Physics}, vol.~56,
  pp.~3661--3681, Sept. 1997.

\bibitem{Boghosian1998}
B.~M. Boghosian and W.~Taylor, ``{Simulating quantum mechanics on a quantum
  computer},'' {\em Physica D-Nonlinear Phenomena}, vol.~120, no.~1,
  pp.~30--42, 1998.

\bibitem{Zalka1998}
C.~Zalka, ``{Efficient Simulation of Quantum Systems by Quantum Computers},''
  {\em Fortschritte der Physik}, vol.~46, no.~6, pp.~877--879, 1998.

\bibitem{Abrams1999}
D.~S. Abrams and S.~Lloyd, ``{Quantum Algorithm Providing Exponential Speed
  Increase for Finding Eigenvalues and Eigenvectors},'' {\em Physical Review
  Letters}, vol.~83, pp.~5162--5165, Dec. 1999.

\bibitem{Berry2007}
D.~W. Berry, G.~Ahokas, R.~Cleve, and B.~C. Sanders, ``{Efficient Quantum
  Algorithms for Simulating Sparse Hamiltonians},'' {\em Communications In
  Mathematical Physics}, vol.~270, pp.~359--371, Mar. 2007.

\bibitem{Kassal2008}
I.~Kassal, S.~P. Jordan, P.~J. Love, M.~Mohseni, and A.~Aspuru-Guzik,
  ``{Polynomial-time quantum algorithm for the simulation of chemical
  dynamics},'' {\em Proceedings of the National Academy of Sciences}, vol.~105,
  pp.~18681--18686, Jan. 2008.

\bibitem{Wiebe2008}
N.~Wiebe, D.~W. Berry, P.~Hoyer, and B.~C. Sanders, ``{Higher Order
  Decompositions of Ordered Operator Exponentials},'' {\em Journal of Physics
  A: Mathematical and Theoretical}, vol.~43, pp.~1--21, Dec. 2010.

\bibitem{Ward2009}
N.~J. Ward, I.~Kassal, and A.~Aspuru-Guzik, ``{Preparation of many-body states
  for quantum simulation},'' {\em Journal Of Chemical Physics}, vol.~130,
  pp.~194105--194114, Dec. 2008.

\bibitem{Sanders2012}
S.~Raeisi, N.~Wiebe, and B.~C. Sanders, ``{Quantum-circuit design for efficient
  simulations of many-body quantum dynamics},'' {\em New Journal Of Physics},
  vol.~14, p.~3017, Oct. 2012.

\bibitem{Sanders2013}
B.~C. Sanders, ``{Efficient Algorithms for Universal Quantum Simulation},''
  {\em Lecture Notes in Computer Science}, vol.~7948, pp.~1--10, July 2013.

\bibitem{Weimer2010}
H.~Weimer, M.~M\"{u}ller, I.~Lesanovsky, P.~Zoller, and H.~P. B\"{u}chler, ``{A
  Rydberg quantum simulator},'' {\em Nature Physics}, vol.~6, pp.~382--388, May
  2010.

\bibitem{Ma2011}
X.-S. Ma, B.~Daki\'{c}, W.~Naylor, A.~Zeilinger, and P.~Walther, ``{Quantum
  simulation of the wavefunction to probe frustrated Heisenberg spin
  systems},'' {\em Nature Physics}, vol.~7, pp.~399--405, May 2011.

\bibitem{Hague2013}
J.~P. Hague, S.~Downes, C.~MacCormick, and P.~E. Kornilovitch, ``{Cold Rydberg
  atoms for quantum simulation of exotic condensed matter interactions},'' {\em
  Journal of Superconductivity and Novel Magnetism}, Oct. 2013.

\bibitem{Cohen2013}
I.~Cohen and A.~Retzker, ``{Quantum Simulation of the Haldane Phase Using
  Trapped Ions},'' {\em e-print arXiv:1310.3715}, Oct. 2013.

\bibitem{Hauke2013}
P.~Hauke, D.~Marcos, M.~Dalmonte, and P.~Zoller, ``{Quantum simulation of a
  lattice Schwinger model in a chain of trapped ions},'' {\em Physical Review
  X}, vol.~3, p.~18, June 2013.

\bibitem{Simon2011}
J.~Simon, W.~S. Bakr, R.~Ma, M.~E. Tai, P.~M. Preiss, and M.~Greiner,
  ``{Quantum simulation of antiferromagnetic spin chains in an optical
  lattice},'' {\em Nature}, vol.~472, no.~7343, pp.~307--312, 2011.

\bibitem{Greiner2009}
J.~I. Gillen, W.~S. Bakr, A.~Peng, P.~Unterwaditzer, S.~F\"{o}lling, and
  M.~Greiner, ``{Two-dimensional quantum gas in a hybrid surface trap},'' {\em
  Physical Review A}, vol.~80, p.~21602, Aug. 2009.

\bibitem{wineland2002}
D.~Leibfried, B.~Demarco, V.~Meyer, M.~Rowe, A.~Ben-Kish, J.~Britton, W.~M.
  Itano, B.~Jelenkovi\'{c}, C.~Langer, T.~Rosenband, and D.~J. Wineland,
  ``{Trapped-Ion Quantum Simulator: Experimental Application to Nonlinear
  Interferometers},'' {\em Physical Review Letters}, vol.~89, p.~247901, Nov.
  2002.

\bibitem{Schaetz2008}
A.~Friedenauer, H.~Schmitz, J.~T. Glueckert, D.~Porras, and T.~Schaetz,
  ``{Simulating a quantum magnet with trapped ions},'' {\em Nature Physics},
  vol.~4, pp.~757--761, Oct. 2008.

\bibitem{Johanning2009}
M.~Johanning, A.~F. Var\'{o}n, and C.~Wunderlich, ``{REVIEW: Quantum
  simulations with cold trapped ions},'' {\em Journal of Physics B: Atomic},
  vol.~42, p.~4009, Aug. 2009.

\bibitem{Ma2012b}
X.-S. Ma, B.~Daki\'{c}, S.~Kropatsche, W.~Naylor, Y.-h. Chan, Z.-x. Gong, L.-m.
  Duan, A.~Zeilinger, and P.~Walther, ``{Photonic quantum simulation of ground
  state configurations of Heisenberg square and checkerboard lattice spin
  systems},'' {\em e-print arXiv:1205.2801}, 2012.

\bibitem{Monroe2013}
P.~Richerme, C.~Senko, J.~Smith, A.~Lee, S.~Korenblit, and C.~Monroe,
  ``{Experimental performance of a quantum simulator: Optimizing adiabatic
  evolution and identifying many-body ground states},'' {\em Physical Review
  A}, vol.~88, p.~12334, July 2013.

\bibitem{Aspuru-Guzik2006}
A.~Aspuru-Guzik, A.~D. Dutoi, P.~J. Love, and M.~Head-Gordon, ``{Simulated
  Quantum Computation of Molecular Energies},'' {\em Science}, vol.~309,
  no.~5741, p.~20, 2006.

\bibitem{Whitfield2010}
J.~D. Whitfield, J.~Biamonte, and A.~Aspuru-Guzik, ``{Simulation of Electronic
  Structure Hamiltonians Using Quantum Computers},'' {\em Molecular Physics},
  vol.~2, no.~2, pp.~106--111, 2010.

\bibitem{Kassal2010}
I.~Kassal, J.~D. Whitfield, A.~Perdomo-Ortiz, M.-H. Yung, and A.~Aspuru-Guzik,
  ``{Simulating chemistry using quantum computers.},'' {\em Annual Review of
  Physical Chemistry}, vol.~62, no.~1, pp.~185--207, 2010.

\bibitem{CodyJones2012}
N.~{Cody Jones}, J.~D. Whitfield, P.~L. McMahon, M.-H. Yung, R.~V. Meter,
  A.~Aspuru-Guzik, and Y.~Yamamoto, ``{Faster quantum chemistry simulation on
  fault-tolerant quantum computers},'' {\em New Journal of Physics}, vol.~14,
  no.~11, pp.~115023(1--35), 2012.

\bibitem{Lanyon2009}
B.~P. Lanyon, J.~D. Whitfield, G.~G. Gillet, M.~E. Goggin, M.~P. Almeida,
  I.~Kassal, J.~D. Biamonte, M.~Mohseni, B.~J. Powell, M.~Barbieri,
  A.~Aspuru-Guzik, and A.~G. White, ``{Towards Quantum Chemistry on a Quantum
  Computer},'' {\em Nature Chemistry}, vol.~2, no.~2, p.~20, 2009.

\bibitem{Wecker2013}
D.~Wecker, B.~Bauer, B.~K. Clark, M.~B. Hastings, and M.~Troyer, ``{Can quantum
  chemistry be performed on a small quantum computer?},'' {\em e-print
  arXiv:1312.1695}, 2013.

\bibitem{Toloui2013}
B.~Toloui and P.~J. Love, ``{Quantum Algorithms for Quantum Chemistry based on
  the sparsity of the CI-matrix},'' {\em e-print arXiv:1312.2579}, 2013.

\bibitem{McClean2013}
A.~Peruzzo, J.~McClean, P.~Shadbolt, M.-H. Yung, X.-Q. Zhou, P.~J. Love,
  A.~Aspuru-Guzik, and J.~L. O'Brien, ``{A variational eigenvalue solver on a
  quantum processor},'' {\em e-print 1304.3061}, 2013.

\bibitem{Welch2013}
J.~Welch, D.~Greenbaum, S.~Mostame, and A.~Aspuru-Guzik, ``{Efficient Quantum
  Circuits for Diagonal Unitaries Without Ancillas},'' {\em e-print
  arXiv:1306.3991}, p.~8, 2013.

\bibitem{Whitfield2013b}
J.~D. Whitfield, ``{Spin-free quantum computational simulations and symmetry
  adapted states},'' {\em Journal of Chemical Physics}, vol.~139, no.~2, 2013.

\bibitem{lu2011}
D.~Lu, N.~Xu, R.~Xu, H.~Chen, J.~Gong, X.~Peng, and J.~Du, ``{Simulation of
  chemical isomerization reaction dynamics on a NMR quantum simulator.},'' {\em
  Physical review letters}, vol.~107, no.~2, p.~020501, 2011.

\bibitem{Yung2013}
M.~H. Yung, J.~Casanova, A.~Mezzacapo, J.~McClean, L.~Lamata, A.~Aspuru-Guzik,
  and E.~Solano, ``{From transistor to trapped-ion computers for quantum
  chemistry},'' {\em e-print arXiv:1307.4326}, p.~9, 2013.

\bibitem{Seeley2012}
J.~T. Seeley, M.~J. Richard, and P.~J. Love, ``{The Bravyi-Kitaev
  transformation for quantum computation of electronic structure},'' {\em
  Journal of Chemical Physics}, vol.~137, no.~22, 2012.

\bibitem{Bravyi2000}
S.~Bravyi and A.~Kitaev, ``{Fermionic quantum computation},'' {\em Annals of
  Physics}, vol.~298, no.~1, p.~18, 2000.

\bibitem{Jordan1928}
P.~Jordan and E.~Wigner, ``\"{u}ber das paulische \"{a}quivalenzverbot,'' {\em
  Zeitschrift f\"{u}r Physik}, vol.~47, no.~9-10, pp.~631--651, 1928.

\bibitem{Somma2002}
R.~Somma, G.~Ortiz, J.~Gubernatis, E.~Knill, and R.~Laflamme, ``{Simulating
  physical phenomena by quantum networks},'' {\em Physical Review A}, vol.~65,
  no.~4, p.~17, 2002.

\bibitem{Kempe2004}
J.~Kempe, A.~Kitaev, and O.~Regev, ``{The Complexity of the Local Hamiltonian
  Problem},'' {\em SIAM Journal on Computing}, vol.~35, no.~5, p.~30, 2004.

\bibitem{Jordan2008}
S.~P. Jordan and E.~Farhi, ``{Perturbative Gadgets at Arbitrary Orders},'' {\em
  Physical Review A}, vol.~77, no.~6, pp.~1--8, 2008.

\bibitem{Oliveira2005}
R.~Oliveira and B.~M. Terhal, ``{The complexity of quantum spin systems on a
  two-dimensional square lattice},'' {\em Quant Inf Comp}, vol.~8, no.~10,
  p.~19, 2005.

\bibitem{Cao2013}
Y.~Cao, R.~Babbush, J.~Biamonte, and S.~Kais, ``{Experimentally Realizable
  Hamiltonian Gadgets},'' {\em e-print arXiv:1311.2555}, p.~43, 2013.

\bibitem{Wang2008}
H.~Wang, S.~Kais, A.~Aspuru-Guzik, and M.~R. Hoffmann, ``{Quantum algorithm for
  obtaining the energy spectrum of molecular systems.},'' {\em Physical
  chemistry chemical physics : PCCP}, vol.~10, no.~35, pp.~5388--5393, 2008.

\bibitem{Veis2010}
L.~Veis and J.~Pittner, ``{Quantum computing applied to calculations of
  molecular energies: CH2 benchmark.},'' {\em The Journal of chemical physics},
  vol.~133, no.~19, p.~194106, 2010.

\bibitem{Li2011}
Z.~Li, M.-H. Yung, H.~Chen, D.~Lu, J.~D. Whitfield, X.~Peng, A.~Aspuru-Guzik,
  and J.~Du, ``{Solving Quantum Ground-State Problems with Nuclear Magnetic
  Resonance},'' {\em Scientific Reports}, vol.~1, no.~88, 2011.

\bibitem{Du2010}
J.~Du, N.~Xu, X.~Peng, P.~Wang, S.~Wu, and D.~Lu, ``{NMR implementation of a
  molecular hydrogen quantum simulation with adiabatic state preparation.},''
  {\em Physical review letters}, vol.~104, no.~3, p.~030502, 2010.

\bibitem{Schuch2009}
N.~Schuch and F.~Verstraete, ``{Computational complexity of interacting
  electrons and fundamental limitations of density functional theory},'' {\em
  Nature Physics}, vol.~5, no.~10, pp.~732--735, 2009.

\bibitem{Whitfield2013}
J.~D. Whitfield, P.~J. Love, and A.~Aspuru-Guzik, ``{Computational complexity
  in electronic structure},'' {\em Physical Chemistry Chemical Physics},
  vol.~15, no.~2, pp.~397--411, 2013.

\bibitem{Bravyi2008}
S.~Bravyi, D.~P. DiVincenzo, D.~Loss, and B.~M. Terhal, ``{Quantum simulation
  of many-body Hamiltonians using perturbation theory with bounded-strength
  interactions.},'' {\em Physical review letters}, vol.~101, no.~7, p.~070503,
  2008.

\bibitem{Biamonte2007}
J.~D. Biamonte and P.~J. Love, ``{Realizable Hamiltonians for Universal
  Adiabatic Quantum Computers},'' {\em Physical Review A}, vol.~78, no.~1,
  pp.~1--7, 2007.

\bibitem{Biamonte2008}
J.~D. Biamonte, ``{Non-perturbative k-body to two-body commuting conversion
  Hamiltonians and embedding problem instances into Ising spins},'' {\em
  Physical Review A}, vol.~77, no.~5, pp.~1--8, 2008.

\bibitem{Duan2011}
Q.-H. Duan and P.-X. Chen, ``{Realization of Universal Adiabatic Quantum
  Computation with Fewer Physical Resources},'' {\em Physical Review A},
  vol.~84, no.~4, p.~4, 2011.

\bibitem{Babbush2013b}
R.~Babbush, B.~O'Gorman, and A.~A. Aspuru-Guzik, ``{Resource efficient gadgets
  for compiling adiabatic quantum optimization problems},'' {\em Annalen der
  Physik}, vol.~525, no.~10-11, pp.~877--888, 2013.

\bibitem{Nagaj2007}
D.~Nagaj and S.~Mozes, ``{New construction for a QMA complete three-local
  Hamiltonian},'' {\em Journal of Mathematical Physics}, vol.~48, p.~2104, July
  2007.

\bibitem{Nagaj2010}
D.~Nagaj, ``{Fast universal quantum computation with railroad-switch local
  Hamiltonians},'' {\em Journal of Mathematical Physics}, vol.~51, p.~2201,
  June 2010.

\bibitem{Gosset2013}
D.~Gosset and D.~Nagaj, ``{Quantum 3-SAT is QMA1-complete},'' {\em e-print
  arXiv:1302.0290}, pp.~1--44, Feb. 2013.

\bibitem{Childs2013}
A.~M. Childs, D.~Gosset, and Z.~Webb, ``{The Bose-Hubbard model is
  QMA-complete},'' {\em e-print arXiv:1311.3297}.

\bibitem{Verstraete2005}
F.~Verstraete and J.~I. Cirac, ``{Mapping local Hamiltonians of fermions to
  local Hamiltonians of spins},'' {\em Journal of Statistical Mechanics: Theory
  and Experiment}, p.~P09012, Aug. 2005.

\bibitem{Bravyi2006}
S.~Bravyi, D.~P. DiVincenzo, R.~I. Oliveira, and B.~M. Terhal, ``{The
  Complexity of Stoquastic Local Hamiltonian Problems},'' {\em Quantum
  Information \& Computation}, vol.~8, no.~5, pp.~361--385, 2008.

\end{thebibliography}

\section*{Appendix}

In this Appendix we provide details of some of the calculations of terms in the perturbative expansion of the self-energy, for the case $k=3$. The projector onto the ancilla ground space is:
\begin{equation}
\Pi_-=\ket{000}\bra{000}+\ket{111}\bra{111}.
\end{equation}
The projector onto the high energy subspace is simply the projector onto the orthogonal complement of the ground space:
\begin{equation}
\Pi_+=\openone-\Pi_-.
\end{equation}
In general we may write our perturbation as follows:
\begin{equation}
V=(H_{\rm else}+\Lambda)\otimes \openone + \mu\sum_{i\in\{a,b,c\}} O_i\otimes X_i
\end{equation}
We first calculate $V_-$:
\begin{eqnarray}
V_- & = &\Pi_- V \,\Pi_-\\
& = &\openone\otimes\Pi_- \left[(H_{\rm else}+\Lambda)\otimes \openone\right]\openone\otimes\Pi_- + \mu\openone\otimes\Pi_-\left[\sum_{i\in\{a,b,c\}} O_i\otimes X_i\right]\openone\otimes\Pi_-\nonumber\\
& = & (H_{\rm else}+\Lambda)\otimes \openone\otimes\Pi_- + \mu\sum_{i\in\{a,b,c\}} O_i\otimes \Pi_-X_i\,\Pi_-\nonumber\\
& = & (H_{\rm else}+\Lambda)\otimes \openone\otimes\Pi_-\nonumber
\end{eqnarray}
where we used $\Pi_-X_i\,\Pi_-=0~\forall i$ in the last line.  Next we calculate $V_+$, and to do so we exploit the orthogonality of the low and high energy subspaces:
\begin{equation}
V_+ = (\openone-\Pi_-)V(\openone-\Pi_-) = V-\Pi_- V-V\,\Pi_- +\Pi_-V\,\Pi_- = V+V_--\Pi_- V-V\,\Pi_-.
\end{equation}
Now we compute term by term:
\begin{eqnarray}
\Pi_- V & = & (H_{\rm else}+\Lambda)\otimes\Pi_- +\mu\sum O_i \Pi_-X_i\\
V\Pi_-  & = & (H_{\rm else}+\Lambda)\otimes\Pi_- +\mu\sum O_i X_i\, \Pi_-.
\end{eqnarray}
We obtain:
\begin{eqnarray}
V_+ & = & (H_{\rm else}+\Lambda)\otimes(1-\Pi_--\Pi_-+\Pi_-) + \mu\sum O_i\otimes[X_i+\Pi_-X_i\Pi_--\Pi_-X_i-X_i\,\Pi_-]\\
& = & (H_{\rm else}+\Lambda)\otimes\Pi_+ + \mu\sum O_i\otimes[X_i-\Pi_-X_i-X_i\,\Pi_-]\nonumber
\end{eqnarray}
This gives, setting $O_a=A$, $O_b=B$, $O_c=C$:
\begin{eqnarray}
V_+&=&(H_{\rm else}+\Lambda)\otimes\Pi_+\\
&+&\mu A\otimes \left(\ket{001}\bra{101}+\ket{010}\bra{110}+\ket{101}\bra{001}+\ket{110}\bra{010}\right)\nonumber\\
&+&\mu B\otimes \left(\ket{001}\bra{011}+\ket{011}\bra{001}+\ket{100}\bra{110}+\ket{110}\bra{100}\right)\nonumber\\
&+&\mu C\otimes \left(\ket{010}\bra{011}+\ket{100}\bra{101}+\ket{011}\bra{010}+\ket{101}\bra{100}\right)\nonumber.
\end{eqnarray}
Next we calculate $V_{-+}=\Pi_- V\, \Pi_+$:
\begin{eqnarray}
V_{-+} & = & \Pi_-V(\openone-\Pi_-) = \Pi_- V -V_-\\
& = &(H_{\rm else}+\Lambda)\otimes\Pi_- +\mu\sum O_i\otimes \Pi_-X_i-(H_{\rm else}+\Lambda)\otimes\Pi_- = \mu\sum O_i\otimes \Pi_-X_i\nonumber.
\end{eqnarray}
This gives, setting $O_a=A$, $O_b=B$, $O_c=C$:
\begin{eqnarray}
V_{-+}&=&\mu A\otimes \left(\ket{100}\bra{000}+\ket{011}\bra{111}\right)\\
&+&\mu B\otimes \left(\ket{010}\bra{000}+\ket{101}\bra{111}\right)\nonumber\\
&+&\mu C\otimes \left(\ket{001}\bra{000}+\ket{110}\bra{111}\right)\nonumber
\end{eqnarray}
and $V_{+-}=\Pi_+ V\, \Pi_-$:
\begin{eqnarray}
V_{+-} & = & (\openone-\Pi_-)V\,\Pi_- = V\,\Pi_-  -V_-\\
&=&(H_{\rm else}+\Lambda)\otimes\Pi_- +\mu\sum O_i \otimes X_i\,\Pi_--(H_{\rm else}+\Lambda)\otimes\Pi_- = \mu\sum O_i\otimes X_i\,\Pi_-\nonumber.
\end{eqnarray}
This gives, setting $O_a=A$, $O_b=B$, $O_c=C$:
\begin{eqnarray}
V_{+-}&=&\mu O_A\otimes \left(\ket{000}\bra{100}+\ket{111}\bra{011}\right)\\
&+&\mu O_B\otimes \left(\ket{000}\bra{010}+\ket{111}\bra{101}\right)\nonumber\\
&+&\mu O_C\otimes \left(\ket{000}\bra{001}+\ket{111}\bra{110}\right)\nonumber.
\end{eqnarray}

The first term in the perturbation series is $V_{-+}V_+V_{+-}$. This is best computed using the expressions for these quantities in terms of the projectors, rather than the explicit expressions in the logical basis:
\begin{eqnarray}
V_{-+}V_+V_{+-}&=&\mu^3\sum O_i\otimes \Pi_-X_i\biggl((H_{\rm else}+\Lambda)\otimes\Pi_+\biggr)\sum_k O_k \otimes X_k\,\Pi_-\\
&+& \mu^3\sum O_i\otimes \Pi_-X_i\biggl(\mu\sum_j O_j\otimes[X_j-\Pi_-X_j-X_j\,\Pi_-]\biggr)\sum_k O_k\otimes X_k\,\Pi_-\nonumber
\end{eqnarray}
using the orthogonality of the high and low subspaces:
\begin{eqnarray}
V_{-+}V_+V_{+-}&=& \mu\sum O_i \otimes\Pi_-X_i\biggl(\mu\sum_j O_j\otimes[X_j-\Pi_-X_j-X_j\,\Pi_-]\biggr)\mu\sum_k O_k \otimes X_k\,\Pi_-\\
&=& \mu^3\sum_{ijk} O_iO_jO_k\otimes \biggl(\Pi_-X_i[X_j-\Pi_-X_j-X_j\,\Pi_-]  X_k\,\Pi_-\biggr)\nonumber\\
&=& \mu^3\sum_{ijk} O_iO_jO_k\otimes \biggl(\Pi_-X_i[X_jX_k\,\Pi_--  \Pi_-X_jX_k\,\Pi_-  -  X_j\Pi_-X_k\,\Pi_-   ]  \biggr)\nonumber\\
&=& \mu^3\sum_{ijk} O_iO_jO_k\otimes \biggl(\Pi_-X_i[X_jX_k\,\Pi_--  \Pi_-X_jX_k\,\Pi_- ]  \biggr)\nonumber\\
&=& \mu^3\sum_{ijk} O_iO_jO_k\otimes \biggl(\Pi_-X_iX_jX_k\,\Pi_-- \Pi_-X_i \Pi_-X_jX_k\,\Pi_-   \biggr)\nonumber\\
&=& \mu^3\sum_{ijk} O_iO_jO_k\otimes \biggl(\Pi_-X_iX_jX_k\,\Pi_-\biggr)\nonumber
\end{eqnarray}
where we have repeatedly used $\Pi_-X_i\,\Pi_-=0
\quad\forall i$.

\end{document}